\documentclass[sigconf,nonacm]{acmart}

\AtBeginDocument{%
  }

\setcopyright{none}
\usepackage{graphicx}
\usepackage{caption}
\usepackage{subcaption}
\usepackage{algorithm}
\usepackage{algpseudocode}
\usepackage{amsmath}
\usepackage{bbm}
\usepackage{amsthm}
\usepackage{amsfonts}
\usepackage{url}
\usepackage{xspace}
\usepackage{multirow}
\usepackage{booktabs}
\usepackage{mathtools}
\usepackage{color}

\usepackage{multirow}
\usepackage{bbding}
\usepackage{colortbl}
\usepackage{textcomp}
\usepackage{comment}
\usepackage{enumerate}
\usepackage{tcolorbox}
\tcbuselibrary{breakable,skins}
\usepackage{pifont}

\usepackage{enumitem}
\usepackage{tabularx}

\usepackage{array}
\usepackage{ragged2e}
\usepackage{longtable}
\usepackage{hyperref}

\setlist[itemize]{leftmargin=*}

\usepackage{xcolor}
\hypersetup{
  colorlinks,
  linkcolor={blue!70!green},
  citecolor={green!70!blue},
  urlcolor={orange!70!red}
}

\usepackage{hyperref}
\makeatletter
\def\UrlAlphabet{%
      \do\a\do\b\do\c\do\d\do\e\do\f\do\g\do\h\do\i\do\j%
      \do\k\do\l\do\m\do\n\do\o\do\p\do\q\do\r\do\s\do\t%
      \do\u\do\v\do\w\do\x\do\y\do\z\do\A\do\B\do\C\do\D%
      \do\E\do\F\do\G\do\H\do\I\do\J\do\K\do\L\do\M\do\N%
      \do\O\do\P\do\Q\do\R\do\S\do\T\do\U\do\V\do\W\do\X%
      \do\Y\do\Z}
\def\UrlDigits{\do\1\do\2\do\3\do\4\do\5\do\6\do\7\do\8\do\9\do\0}
\g@addto@macro{\UrlBreaks}{\UrlOrds}
\g@addto@macro{\UrlBreaks}{\UrlAlphabet}
\g@addto@macro{\UrlBreaks}{\UrlDigits}
\makeatother

\definecolor{mygray}{gray}{.9}

\newcommand{\ie}{\textit{i.e.}\xspace}

\newcommand{\mypara}[1]{\noindent\textbf{#1.} \xspace}

\newcommand{\sysname}{MarkSec\xspace}

\definecolor{revision}{RGB}{50,80,255}

\newcommand{\revstart}{\begin{color}{revision}}
\newcommand{\revend}{~\!\!\end{color}}

\definecolor{takeawaybg}{RGB}{244,249,253}
\definecolor{takeawayframe}{RGB}{157,183,205}
\definecolor{takeawayaccent}{RGB}{58,114,173}
\definecolor{takeawayrule}{RGB}{191,212,230}

\newcommand{\takeawaylineinset}{1.6mm}
\newcommand{\takeawayleftx}{1.45mm}
\newcommand{\takeawayrightx}{1.25mm}
\newcommand{\takeawaydecorunbroken}{%
  \draw[takeawayaccent, line width=2.8pt, line cap=round]
    ([xshift=\takeawayleftx, yshift=-\takeawaylineinset]frame.north west) --
    ([xshift=\takeawayleftx, yshift=\takeawaylineinset]frame.south west);
  \draw[takeawayframe, line width=1.15pt, line cap=round]
    ([xshift=-\takeawayrightx, yshift=-\takeawaylineinset]frame.north east) --
    ([xshift=-\takeawayrightx, yshift=\takeawaylineinset]frame.south east);
}
\newcommand{\takeawaydecorfirst}{%
  \draw[takeawayaccent, line width=2.8pt, line cap=round]
    ([xshift=\takeawayleftx, yshift=-\takeawaylineinset]frame.north west) --
    ([xshift=\takeawayleftx]frame.south west);
  \draw[takeawayframe, line width=1.15pt, line cap=round]
    ([xshift=-\takeawayrightx, yshift=-\takeawaylineinset]frame.north east) --
    ([xshift=-\takeawayrightx]frame.south east);
}
\newcommand{\takeawaydecormiddle}{%
  \draw[takeawayaccent, line width=2.8pt]
    ([xshift=\takeawayleftx]frame.north west) --
    ([xshift=\takeawayleftx]frame.south west);
  \draw[takeawayframe, line width=1.15pt]
    ([xshift=-\takeawayrightx]frame.north east) --
    ([xshift=-\takeawayrightx]frame.south east);
}
\newcommand{\takeawaydecorlast}{%
  \draw[takeawayaccent, line width=2.8pt, line cap=round]
    ([xshift=\takeawayleftx]frame.north west) --
    ([xshift=\takeawayleftx, yshift=\takeawaylineinset]frame.south west);
  \draw[takeawayframe, line width=1.15pt, line cap=round]
    ([xshift=-\takeawayrightx]frame.north east) --
    ([xshift=-\takeawayrightx, yshift=\takeawaylineinset]frame.south east);
}

\newtcolorbox{takeawaypanel}{
  enhanced,
  breakable=false,
  colback=takeawaybg,
  colframe=takeawayframe,
  boxrule=0.75pt,
  arc=2.2mm,
  left=3.2mm,
  right=2.8mm,
  top=1.25mm,
  bottom=1.15mm,
  overlay unbroken={\takeawaydecorunbroken},
  overlay first={\takeawaydecorfirst},
  overlay middle={\takeawaydecormiddle},
  overlay last={\takeawaydecorlast},
  before skip=6pt plus 2pt minus 1pt,
  after skip=8pt plus 2pt minus 1pt
}
\newcounter{findingbox}

\newcommand{\takeaway}[1]{%
  \refstepcounter{findingbox}%
  \begin{takeawaypanel}%
  \noindent\textbf{\textcolor{takeawayaccent}{Finding~\thefindingbox.}}\enspace #1%
  \end{takeawaypanel}%
}

\definecolor{new content}{RGB}{0,255,0}

\usepackage{etoolbox}
\makeatletter
\patchcmd{\hyper@makecurrent}{%
    \ifx\Hy@param\Hy@chapterstring
        \let\Hy@param\Hy@chapapp
    \fi
}{%
    \iftoggle{inappendix}{
        \@checkappendixparam{chapter}%
        \@checkappendixparam{section}%
        \@checkappendixparam{subsection}%
        \@checkappendixparam{subsubsection}%
        \@checkappendixparam{paragraph}%
        \@checkappendixparam{subparagraph}%
    }{}%
}{}{\errmessage{failed to patch}}

\newcommand*{\@checkappendixparam}[1]{%
    \def\@checkappendixparamtmp{#1}
    \ifx\Hy@param\@checkappendixparamtmp
        \let\Hy@param\Hy@appendixstring
    \fi
}
\makeatletter

\newtoggle{inappendix}
\togglefalse{inappendix}

\apptocmd{\appendix}{\toggletrue{inappendix}}{}{\errmessage{failed to patch}}

\usepackage{tikz}
\usepackage{oplotsymbl}

\usepackage[colorinlistoftodos]{todonotes}

\newfontface\marksecSymbolFont{DejaVuSans.ttf}[Path=fonts/]
\newcommand{\cmark}{{\marksecSymbolFont\symbol{"2713}}}
\newcommand{\xmark}{{\marksecSymbolFont\symbol{"2717}}}

\newcommand{\KGW}{\textsf{KGW}\xspace}
\newcommand{\Unigram}{\textsf{Unigram}\xspace}
\newcommand{\SIR}{\textsf{SIR}\xspace}
\newcommand{\SynthID}{\textsf{SynthID}\xspace}
\newcommand{\UW}{\textsf{UW}\xspace}
\newcommand{\DIP}{\textsf{DIP}\xspace}

\newcommand{\DIPPER}{\textsf{DIPPER}\xspace}
\newcommand{\LLMP}{\textsf{LLMP}\xspace}
\newcommand{\SIRA}{\textsf{SIRA}\xspace}
\newcommand{\RW}{\textsf{RW}\xspace}
\newcommand{\SA}{\textsf{SA}\xspace}
\newcommand{\SCTS}{\textsf{SCTS}\xspace}
\newcommand{\JSV}{\textsf{JSV}\xspace}
\newcommand{\Bfour}{\textsf{B4}\xspace}
\newcommand{\MIP}{\textsf{MIP}\xspace}
\newcommand{\DEMARK}{\textsf{DE-MARK}\xspace}

\newcommand{\SCTSSt}{\textsf{SCTS-St}\xspace}
\newcommand{\SCTSSc}{\textsf{SCTS-Sc}\xspace}

\newcommand{\JSVSt}{\textsf{JSV-St}\xspace}
\newcommand{\JSVSc}{\textsf{JSV-Sc}\xspace}
\newcommand{\JSVSp}{\textsf{JSV-Sp}\xspace}

\newcommand{\BfourSt}{\textsf{B4-St}\xspace}
\newcommand{\BfourSc}{\textsf{B4-Sc}\xspace}

\newcommand{\MIPSt}{\textsf{MIP-St}\xspace}
\newcommand{\MIPSc}{\textsf{MIP-Sc}\xspace}

\newcommand{\DEMARKSt}{\textsf{DE-MARK-St}\xspace}
\newcommand{\DEMARKSc}{\textsf{DE-MARK-Sc}\xspace}
\newcommand{\DEMARKSp}{\textsf{DE-MARK-Sp}\xspace}

\makeatother
\newcommand{\ASRobs}{\ensuremath{\mathrm{ASR}}}

\newcommand{\indicator}{\mathbf{1}}

\begin{document}
\raggedbottom

\title{\sysname: Capability-Aware Evaluation of Adversarial Attacks Against LLM Watermarks}

\input{authors}
\renewcommand{\shortauthors}{Li et al.}


\begin{abstract}
LLM watermarking has emerged as a promising solution for tracing the origin of LLM-generated content to help limit accidental or deliberate misuse.
It faces various adversarial attacks: stealing, which recovers watermark-related information; scrubbing, which removes watermark signals from marked text; and spoofing, which forges text that a detector accepts as watermarked without actual watermark embedding.
However, these attacks are often studied in isolation, leaving their connections unclear.
Current evaluations also often lack shared detector calibration, metric definitions, and reporting semantics, making it difficult to compare the risks faced by different watermarks.
In addition, attack success and text quality are measured separately, making it difficult to identify attacks that are both effective and quality-preserving.

To this end, we propose \sysname, a general framework that unifies analyses of stealing, scrubbing, and spoofing.
Building on this framework, we evaluate different attacks under a unified reporting protocol.
We introduce a quality-constrained attack success metric to assess attack effectiveness and text quality jointly and support comparisons across watermark methods.
Evaluations across representative watermark families, attack methods, LLMs, and datasets reveal three key findings.
First, attacks that appear strongest by watermark removal alone can fall behind general rewriting under a criterion that jointly measures success and text quality.
Second, general rewriting remains a strong baseline across watermark families, while its advantage over other scrubbing attacks varies by family.
Third, in a case study of one watermark family, stealing-based scrubbers often underperform the best general-scrubbing baselines when text quality is required.
These results show that apparent attack winners depend on text-quality constraints, attack generality, and capability assumptions.
\end{abstract}

\ccsdesc[500]{Security and privacy~Software and application security}
\ccsdesc[500]{Computing methodologies~Natural language processing}


\keywords{LLM Watermarks, Watermark Security, LLM Content Provenance} 


\maketitle

\section{Introduction}

\textit{Large language models} (LLMs) are increasingly applied in software development~\cite{peng2023impact,jiang2026survey}, education~\cite{xu2024leducation,wang2024leducation}, and online content production~\cite{wang2024weaver,liang2025widespread}.
As model outputs become more fluent and human-like, distinguishing LLM-generated text from human-written text becomes more difficult~\cite{weidinger2022taxonomy,abdelnabi2023signed}, which creates provenance and misuse concerns.
For instance, schools and universities have debated or restricted ChatGPT use because of concerns about plagiarism and academic dishonesty~\cite{ap2023chatgptschools,abc2023chatgptschools}.
AI-generated spam and low-quality synthetic content are increasingly polluting online platforms and search results~\cite{abc2024aiflood}.
These trends make tracing LLM-generated content increasingly important in real-world LLM deployments~\cite{dathathri2024scalable}.
\textit{LLM watermarking} has therefore emerged as a promising solution because it embeds machine-verifiable signals during generation~\cite{kirchenbauer2023watermark} and does not depend only on post-hoc detection, which is fragile under paraphrasing~\cite{mitchell2023detectgpt,krishna2023paraphrasing}.

LLM watermarking is useful only if it remains reliable and secure under adversarial manipulation.
In practice, it faces three main threat types.
\textit{Stealing} attacks recover watermark-related information, such as key-dependent token preferences or detector scoring rules, and can use it to support stronger downstream attacks~\cite{jovanovic2024watermark}.
\textit{Scrubbing} attacks remove or weaken watermark signals in marked text through paraphrasing, rewriting, or targeted editing~\cite{sadasivan2023can,cheng2025revealing}.
\textit{Spoofing} attacks forge watermark-like outputs without actual watermark embedding, potentially creating false tracing signals~\cite{chen2024demark}.

Despite the broad attack surfaces, existing studies investigate these attacks mostly in isolation.
Furthermore, these attacks are rarely evaluated with shared detector calibration, metric definitions, and reporting semantics.
Prior studies use different models, watermark families, datasets, attack budgets, and reporting protocols, making it difficult to compare attack results across studies~\cite{tu2024waterbench,liang2025watermark}.
In addition, the attack success rate and text quality are often measured separately, making it hard to identify attacks that are both effective and quality-preserving.
As a result, it remains unclear how these attacks relate to one another, which attacks act as broad baselines versus narrow specialists, and which vulnerabilities matter most in practice.

\mypara{Unified Framework}
To address these issues, we study \textit{stealing}, \textit{scrubbing}, and \textit{spoofing} as a connected attack space.
Our key observation is that the three attack types are connected by how watermark-related information is obtained and reused.
In particular, \textit{stealing} can recover reusable artifacts, such as token preferences, detector statistics, or proxy distributions.
The same artifacts can then support downstream \textit{scrubbing}, which removes watermark evidence from existing text, or \textit{spoofing}, which creates false attribution by generating watermark-like text.
Motivated by this observation, we introduce a unified taxonomy that organizes these attacks by objective, capability, recovered artifact, and downstream use.

\mypara{Experimental Evaluation}
Building on this taxonomy, we evaluate adversarial attacks against LLM watermarks under a unified benchmark protocol.
Our evaluation covers representative watermark families, attack methods, datasets, and LLMs.
We report quality-constrained attack success, QSR, which jointly counts recorded removal and acceptance by a declared automatic quality gate.
The main general-scrubbing benchmark focuses on three questions: how robust watermarks are overall, whether attack strength is broad or family-conditioned, and how rankings change under stricter Prometheus quality gates.
Two further analyses examine Base--Instruct \KGW differences across prompting protocols and the use of recovered watermark information for targeted scrubbing or spoofing.
We also analyze quality-gate sensitivity across all six watermark families on common Llama/Qwen support.

\mypara{Main Findings}
Our evaluation shows that the reported attack ordering depends on the success definition and the automatic quality gate.
On C4, \SIRA leads raw removal, but \LLMP leads quality-aware scrubbing; \DIPPER remains the closest quality-aware alternative in several cells.
Watermark removal alone can therefore overstate practical attack strength.
Family-conditioned analysis further shows that \LLMP is the strongest quality-aware general-scrubbing baseline on C4, with the narrowest margins under \UW and \KGW.
\LLMP achieves the highest mean QSR on C4 at all three Prometheus gates.
The Instruct--Base QSR gap is smaller under wrapper-disabled prompting than under native prompting in the evaluated \KGW settings.
Among the stealing-based scrubbers, \Bfour leads QSR@3 in all four \KGW settings; in spoofing, \DEMARKSp reaches 98--100\% detector acceptance and 47--64\% QSSR@3.
The results call for a capability-aware reading: raw removal, quality-aware success, and recovered-artifact assumptions must be reported together.
These findings show that apparent attack winners depend on text-quality constraints, watermark family, deployment format, and attacker capability assumptions.

\mypara{\sysname}
We implement \sysname as a modular and reusable tool for evaluating adversarial attacks against LLM watermarks.
\sysname provides common interfaces for watermark generation, attack execution, detector calibration, text-quality evaluation, and result reporting.
It supports the joint study of \textit{stealing}, \textit{scrubbing}, and \textit{spoofing} while keeping their capability assumptions explicit.
The reporting pipeline groups general scrubbing, stealing-based scrubbing, and spoofing results by their access assumptions and downstream objectives.

\section{Preliminaries}

\subsection{LLM Text Generation}
\label{ssec:llm_generation}

Most modern text-generation LLMs are Transformer-based autoregressive models.
Given an input prompt, the LLM first predicts a probability distribution over its vocabulary, which consists of all possible tokens.
It then selects one token from this distribution and appends it to the current text.
This process repeats until the model generates an end-of-sequence token or reaches a maximum length.


Formally, we view an LLM as a probabilistic generator parameterized by $\theta$.
Let $\mathcal{V}$ denote the token vocabulary, where $|\mathcal{V}|$ is the vocabulary size.
Given a prompt sequence $\mathbf{x}^{(p)} = \{x_1, \dots, x_M\}$, the model generates a response sequence $\mathbf{x}^{(r)} = \{x_{M+1}, \dots, x_{M+T}\}$ token by token.

At each generation step $t$, the model takes the prompt and all previously generated tokens $\mathbf{x}_{<t} = \{x_1, \dots, x_{t-1}\}$ as input.
It computes a logit vector $\mathbf{l}_t \in \mathbb{R}^{|\mathcal{V}|}$, where each entry gives an unnormalized score for one token in the vocabulary.
The logits are then normalized by a softmax function to produce a probability distribution $P_t$ over the vocabulary:
\begin{equation}
    P_t(v \mid \mathbf{x}_{<t}) =
    \frac{\exp(\mathbf{l}_t[v])}
    {\sum_{v' \in \mathcal{V}} \exp(\mathbf{l}_t[v'])},
    \quad v \in \mathcal{V}.
\end{equation}
The next token $x_t$ is selected from this distribution using a decoding strategy such as multinomial sampling, greedy decoding, or beam search.

\subsection{LLM Watermarks}
\label{ssec:watermark}

An LLM watermark is a machine-verifiable signal embedded into LLM-generated text for later provenance verification.
A watermarked generation process should produce text that remains natural to human readers while carrying detectable statistical evidence.
In general, LLM watermarks can be categorized by when and where the watermark signal is introduced.
\textit{Model-parameter watermarks} embed the signal into model-side parameters or learned watermark components, for example, by modifying parameters, fine-tuning behavior, or trigger-response patterns~\cite{liu2023unforgeable,liu2023semantic}.
\textit{Inference-time watermarks} embed the signal during decoding by modifying logits, token probabilities, or the sampling rule without changing model parameters~\cite{kirchenbauer2023watermark,dathathri2024scalable}.
\textit{Post-hoc watermarks} introduce the signal after ordinary text generation, typically by editing, selecting, or filtering outputs according to a watermark rule~\cite{lau2024waterfall}.



In this paper, we focus on \textit{inference-time} LLM watermarks.
As described in~\autoref{ssec:llm_generation}, an LLM first produces token logits, then converts them into a probability distribution (\ie, posterior), and finally samples the next token.
Therefore, inference-time LLM watermarks can be further classified into three types according to where they intervene in this pipeline: logits-based, posterior-based, and sampling-based.

\mypara{Logits-Based Watermarks}
The general idea is to modify the model logits so that some key-dependent tokens become more likely to be sampled.
At generation step $t$, a key-dependent rule assigns a bias to each candidate token, and the watermarked logits can be written as
\begin{equation}
    \mathbf{l}^{W}_t[v] = \mathbf{l}_t[v] + b_{\xi,t}(v),
\end{equation}
where $\mathbf{l}_t[v]$ is the original logit of token $v$, $\xi$ is the watermark key, and $b_{\xi,t}(v)$ is a key-dependent bias.
After softmax, tokens with larger positive bias receive higher sampling probability.
The detector then tests whether the generated sequence contains more favored tokens than expected under unwatermarked generation.

\KGW-style watermarks are representative examples of this class.
They use the watermark key and previous context to partition candidate tokens into green and red sets, add a positive bias to green-token logits during decoding, and detect the watermark by testing whether the green-token count is statistically significant~\cite{kirchenbauer2023watermark}.

\mypara{Posterior-Based Watermarks}
Posterior-based watermarks operate on the normalized next-token distribution rather than on the pre-softmax logits.
At generation step $t$, they apply a key-dependent transformation to the original distribution:
\begin{equation}
    P^{W}_t = \mathsf{T}_{\xi,t}(P_t),
\end{equation}
where $P_t$ is the original next-token distribution and $\mathsf{T}_{\xi,t}$ is a key-dependent transformation.
This transformation reallocates probability mass across tokens according to the watermark key while aiming to preserve the overall generation quality.
The detector applies the same key-dependent rule to the observed sequence and tests whether the accumulated token statistics are more consistent with the watermarked distribution than with ordinary generation.

\UW is a representative example of this class.
It constructs a distribution-preserving transformation so that the marginal output distribution remains close to the original model distribution while still leaving a detectable keyed signal~\cite{hu2023unbiased}.

\mypara{Sampling-Based Watermarks}
These methods keep the probability distribution largely unchanged, but use a key-dependent sampling rule to choose the next token.
Formally, a sampling-stage watermark can be written as
\begin{equation}
    x_t = \mathsf{Sample}_{\xi,t}(P_t),
\end{equation}
where $\mathsf{Sample}_{\xi,t}$ denotes a key-dependent sampling rule.
The detector applies the same key-dependent rule to the generated sequence and accumulates a statistic that measures whether the observed tokens are more consistent with the watermarked sampling process than with ordinary sampling.

SynthID~\cite{dathathri2024scalable} is a representative example of this class.
It uses tournament sampling during decoding.
A key-dependent random function scores candidate tokens, and the sampler selects winners through pairwise token competitions, leaving a recoverable statistical signal in the generated sequence~\cite{dathathri2024scalable}.


\mypara{Watermark Detection}
For all the above classes, watermark detection can be abstracted as a key-dependent hypothesis test.
Given a text sequence $\mathbf{x}$ and a watermark key $\xi$, a detector computes a score $S_{\xi}(\mathbf{x})$ that measures how strongly the text matches the expected watermark signal.
The detector then compares this score with a threshold $\tau$:
\begin{equation}
    D_{\xi}(\mathbf{x}) = \indicator[S_{\xi}(\mathbf{x}) \ge \tau].
\end{equation}
Different watermark methods instantiate $S_{\xi}$ differently, such as a green-token count for logits-based schemes, a distributional consistency score for posterior-based schemes, or a key-dependent sampling statistic for sampling-based schemes.
This unified detection view allows us to define attacks and evaluation metrics without depending on one specific watermark design.

\section{Attacks Against LLM Watermarks}
\label{sec:taxonomy}

\subsection{Threat Model}
\label{ssec:threat_models}


\mypara{Attack Objectives}
\textit{Stealing} aims to recover reusable watermark-related information, such as token preferences, counting rules, green-list structure, or a surrogate distribution.
\textit{Scrubbing} aims to transform a watermarked text $y_w$ into an attacked text $y_a$ that is no longer detected as watermarked while preserving text quality.
\textit{Spoofing} aims to generate a forged text $y_s$ that is accepted as watermarked without being produced by the genuine watermarking process.

\mypara{Attacker Access}
We assume the attacker does not know the model parameters $\theta$ or the secret watermark key $\xi$.
The attacker may have black-box access to the watermarked generation API.
Depending on the attack setting, the attacker may also receive the original prompt $p$, the generated watermarked text $y_w$, or victim-side token information such as probabilities, logits, or top-$k$ scores.
These access differences define the attack capabilities and determine which attacks are directly comparable in our evaluation.
We do not assume access to hidden system prompts or proprietary serving templates.
When prompt context is available, it refers to the user-facing original prompt.


\begin{figure*}
    \centering
    \includegraphics[width=0.75\textwidth]{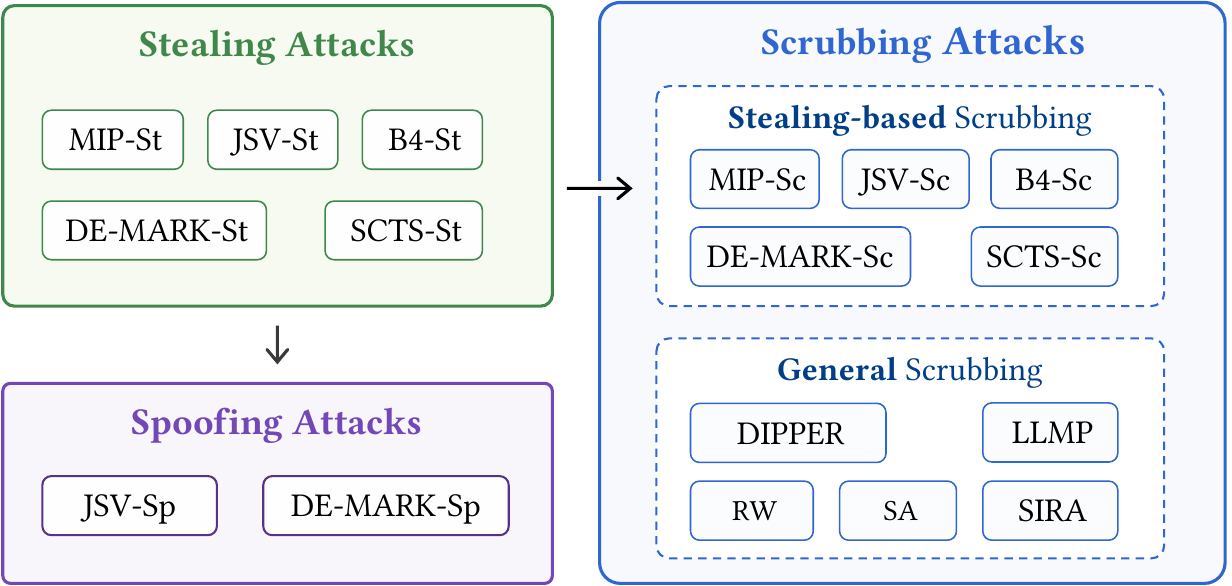}
    \Description{Taxonomy diagram grouping watermark attacks into stealing, scrubbing, and spoofing objectives, with arrows showing how stealing can support downstream spoofing or targeted scrubbing.}
    \caption{Attack taxonomy organized by adversarial objective: \textit{Stealing}, \textit{Scrubbing}, and \textit{Spoofing}.}
    \label{fig:attack_taxonomy}
\end{figure*}

\subsection{Unified Framework}
\label{ssec:unified_framework}

We organize attacks against LLM watermarks by their operational role.
This avoids treating watermark attacks as a flat list of unrelated methods.
Our key observation is that \textit{stealing}, \textit{scrubbing}, and \textit{spoofing} are connected by how watermark-related information is obtained and used.

\textit{Stealing} is an information-recovery objective.
It recovers information that can later support stronger attacks.
For example, a recovered green list, token-color map, or count table can be reused for targeted scrubbing or spoofing, while a distilled proxy distribution is used in this paper only for stealing-based scrubbing.

\textit{Scrubbing} is a removal objective.
It takes watermarked text as input and tries to erase the watermark signal while preserving text quality.
We distinguish between direct scrubbing and stealing-based scrubbing.
Direct scrubbing attacks rewrite, edit, or search over the given watermarked text without first recovering a reusable watermark-related artifact.
Stealing-based scrubbing attacks first recover watermark-related information, such as token preferences or green-list structure, and then use this information to guide removal.

\textit{Spoofing} is a forgery objective.
It tries to create new text that is accepted as watermarked by the detector, even though the text was not produced by the genuine watermark process.
Spoofing therefore attacks the attribution guarantee of the watermark rather than its removal robustness.
In the methods we study, spoofing is usually downstream of stealing or inference because the attacker must recover enough watermark-related information to reproduce the watermark signal and trigger detection.

This relationship is illustrated in \autoref{fig:attack_taxonomy}.
Direct scrubbing attacks form the main comparable removal setting because they operate on the same input type and do not require a prior recovery step.
Stealing-based scrubbing and spoofing are analyzed separately because they rely on additional recovery steps or reusable artifacts.
This separation keeps the main comparison focused while still capturing stronger attack paths in the full threat surface.

The taxonomy also defines the comparison boundary used in our experiments.
\autoref{tab:small_steal_comparison} summarizes stealing-based methods by their stealing source, recovered artifact, and downstream use.
\autoref{tab:direct_comparison} summarizes the direct scrubbing attacks used in the main comparison.
This split keeps general scrubbing results comparable and still covers stronger attack paths that rely on additional capabilities.

\mypara{Role Notation}
Some attack papers support multiple downstream objectives.
To avoid mixing different roles of the same method, we attach an objective suffix when a method appears in more than one part of the taxonomy.
We use \textsc{St} for the stealing role, \textsc{Sc} for the scrubbing role, and \textsc{Sp} for the spoofing role.
For example, \JSVSt denotes the information-recovery stage of \JSV, while \JSVSp denotes its use for spoofing generation.

\subsection{Stealing Attacks}

Stealing attacks recover reusable watermark-related information before downstream use.
The recovered artifact can be explicit, such as a green list or token-color map, or implicit, such as a context count table or proxy distribution.
Its role is to expose information that later supports scrubbing or spoofing.

\begin{table}[t]
\centering
\caption{Required capability of \textit{stealing-based} attacks.}
\label{tab:small_steal_comparison}
    \begin{tabular}{lcccc}
    \toprule
    Method & Online & Prompt & Scores & Spoof \\
    \midrule
    \MIP~\cite{zhang2024stealing}      & \xmark & \xmark & \xmark & \xmark \\
    \SCTS~\cite{wu2024bypassing}       & \cmark & \cmark & \xmark & \xmark \\
    \DEMARK~\cite{chen2024demark}     & \cmark & \cmark & \cmark & \cmark \\
    \JSV~\cite{jovanovic2024watermark} & \xmark & \xmark & \xmark & \cmark \\
    \Bfour~\cite{huang2025b4}             & \xmark & \cmark & \xmark & \xmark \\
    \bottomrule
    \end{tabular}
\end{table}

\mypara{\SCTSSt~\cite{wu2024bypassing}}
Wu and Chandrasekaran propose \SCTS for \KGW-style watermarks, where token choices are biased by a hidden red-green partition.
In its stealing role, \SCTS estimates token-color information through black-box querying.
The attacker queries the watermarked generator with adaptive query prompts and observes which candidate tokens are more likely to appear.
These observations are used to infer a token-color map that approximates the hidden green and red sets.
This recovered color map is the artifact later used by \SCTSSc for targeted substitution.

\mypara{\JSVSt~\cite{jovanovic2024watermark}}
Jovanovi\'c et al. study whether hidden watermark behavior can be approximated from generated samples.
In the stealing role, \JSV builds context-dependent statistics from base and watermarked continuations.
These statistics capture how the watermark shifts token choices under different contexts.
The recovered count table is reusable because it summarizes watermark behavior beyond one attacked sample.
This artifact can later guide \JSVSc for removal or \JSVSp for false attribution.

\mypara{\BfourSt~\cite{huang2025b4}}
Huang et al. formulate \Bfour around a black-box view of the victim watermarked generator.
Its stealing role is to learn a proxy distribution from victim-generated data.
This proxy approximates how the watermarked system allocates probability mass over candidate continuations.
The learned distribution becomes a reusable artifact for downstream rewriting.
\BfourSc then uses this proxy to search for attacked text that reduces the watermark signal while preserving fidelity.

\mypara{\MIPSt~\cite{zhang2024stealing}}
Zhang et al. formulate watermark stealing for green-list-style schemes as a mixed-integer optimization problem.
The attacker uses the collected watermarked and reference text to infer which tokens are likely to belong to the hidden green list.
The optimization stage produces a stolen green-list artifact.
This artifact is reusable because it can guide future token replacement without repeating the full recovery process.
\MIPSc uses the recovered green list for targeted watermark removal.

\mypara{\DEMARKSt~\cite{chen2024demark}}
Chen et al. design \DEMARK to infer watermark rules through token-level querying.
In its stealing role, the attacker sends prompt-guided queries to the victim system and observes how candidate tokens behave under the hidden watermark.
The query process estimates which tokens are favored by the watermark and may also recover parameter-level information.
The resulting query evidence becomes a reusable guide for downstream attack.



\begin{table}[t]
\centering
\caption{Capability questions for \textit{general scrubbing} attacks.}
\label{tab:direct_comparison}
\setlength{\tabcolsep}{5pt}
\renewcommand{\arraystretch}{1.05}
\begin{tabular}{lccccc}
\toprule
Method & Prompt & Scores & Locate & Local & Search \\
\midrule
\DIPPER~\cite{krishna2023paraphrasing} & \xmark & \xmark & \xmark & \xmark & \xmark \\
\LLMP                                  & \xmark & \xmark & \xmark & \xmark & \xmark \\
\SIRA~\cite{cheng2025revealing}        & \xmark & \xmark & \cmark & \cmark & \xmark \\
\RW~\cite{zhang2024watermarks}         & \cmark & \xmark & \xmark & \cmark & \cmark \\
\SA~\cite{chang2025watermark}          & \cmark & \cmark & \cmark & \cmark & \xmark \\
\bottomrule
\end{tabular}
\end{table}

\subsection{Scrubbing Attacks}
\label{ssec:scrubbing_attacks}

Scrubbing attacks form the main general-scrubbing comparison in \sysname.
These attacks take already watermarked text as input and try to remove the watermark signal while preserving text quality.
They differ mainly in what information they use during rewriting, such as the text alone, the original prompt, proxy-model signals, or victim-side scores.

\mypara{\DIPPER~\cite{krishna2023paraphrasing}}
As a generic paraphrasing baseline, \DIPPER rewrites watermarked text into a semantically similar form.
Its intuition is that paraphrasing changes surface tokens and local token patterns, which can disrupt watermark signals embedded in the original wording.
The attack does not use the watermark key, detector feedback, victim-side scores, or a recovered artifact.
In \sysname, \DIPPER represents text-only direct scrubbing.

\mypara{\LLMP}
Our \LLMP baseline uses an LLM to rewrite the full watermarked continuation.
It treats scrubbing as sequence-level rewriting: the model is prompted to preserve the meaning of $y_w$ while producing a new expression.
This attack can preserve high-level semantics because it rewrites the whole continuation at once.
It does not use detector access, victim-side token scores, or a prior stealing stage.
In \sysname, \LLMP represents general LLM-based direct scrubbing.

\mypara{\SIRA~\cite{cheng2025revealing}}
Cheng et al. propose \SIRA to avoid rewriting the entire sequence.
The attack uses self-information from a proxy model to locate spans that appear suspicious under the watermarked text distribution.
It then rewrites only these high-suspicion spans.
This targeted strategy aims to remove watermark evidence while changing fewer parts of the text.
In \sysname, \SIRA represents proxy-guided direct scrubbing.

\mypara{\RW~\cite{zhang2024watermarks}}
\RW performs prompt-aware rewriting through a quality-guided search process.
The attack generates candidate rewrites and keeps candidates that better balance watermark removal and text quality.
The original prompt provides reference context during the search, helping the attack preserve the intended continuation.
This makes \RW different from one-shot paraphrasing baselines.
In \sysname, \RW represents direct scrubbing with prompt-aware search and explicit quality control.

\mypara{\SA~\cite{chang2025watermark}}
Chang et al. propose \SA as a score-aware editing attack.
The attack uses victim-side token information, such as probabilities, logits, or top-$k$ scores, to identify tokens that contribute strongly to the watermark signal.
It edits the text at the token level and searches for alternatives that preserve the local context.
This gives \SA finer control than full-sequence paraphrasing or span-level rewriting.
In \sysname, \SA is a direct scrubber with stronger attack-time access because it uses victim-side token scores.

\mypara{\SCTSSc~\cite{wu2024bypassing}}
The scrubbing role of \SCTS uses the token-color map recovered by \SCTSSt.
Given a watermarked continuation, the attack identifies tokens that are likely to be green under the hidden \KGW-style partition.
It then replaces these tokens with semantically similar alternatives that are less likely to be green.
This directly targets the green-token count used by the detector.
We classify \SCTSSc as stealing-based scrubbing because its substitutions depend on the prior color-recovery stage.

\mypara{\JSVSc~\cite{jovanovic2024watermark}}
\JSVSc uses the context-dependent statistics recovered by \JSVSt to guide removal.
The attack identifies token patterns that are likely to reflect watermarked generation under the recovered statistics.
During rewriting or token selection, it avoids patterns that make the text look watermarked.
This role uses stolen watermark behavior to reduce detector evidence in an existing continuation.
We therefore treat \JSVSc as stealing-based scrubbing.

\mypara{\BfourSc~\cite{huang2025b4}}
\BfourSc uses the proxy distribution learned by \BfourSt to guide rewriting.
The attack searches for continuations that reduce the watermark signal while staying close to the original text under the learned proxy.
Its core trade-off is removal versus fidelity: aggressive changes may evade detection but damage text quality.
The proxy distribution helps the attack navigate this trade-off in a black-box setting.
We classify \BfourSc as stealing-based scrubbing because its removal strategy depends on a reusable learned distribution.

\mypara{\MIPSc~\cite{zhang2024stealing}}
\MIPSc uses the green list recovered by \MIPSt.
The attacker searches for edits that reduce the number of green-list tokens in the watermarked continuation.
Because \KGW-style detection depends on green-token overrepresentation, replacing green tokens with non-green alternatives can reduce the detection score.
This scrubbing role is targeted and watermark-specific.
We classify \MIPSc as stealing-based scrubbing because it depends on the stolen green-list artifact.

\mypara{\DEMARKSc~\cite{chen2024demark}}
\DEMARKSc uses query-derived watermark evidence to remove signals from an existing continuation.
The attack identifies token choices likely favored by the hidden watermark rule.
It then edits the text to avoid these choices while preserving meaning.
This approach uses the recovered rule in the opposite direction of spoofing.
We classify \DEMARKSc as stealing-based scrubbing because its removal strategy is guided by prior stealing.

Direct scrubbing attacks share a common goal: transforming watermarked text into attacked text without recovering a reusable watermark artifact.
They differ in the side information and resources used for rewriting.
All direct scrubbers in our study are detector-free; \autoref{tab:direct_comparison} asks whether an attack uses prompt context, victim-side scores, suspicious span/token localization, local edits, or iterative search.


\subsection{Spoofing Attacks}
\label{ssec:spoofing_attacks}

Spoofing targets false attribution rather than watermark removal.
A spoofing adversary generates text that is accepted by the victim detector as watermarked, even though the text was not produced by the actual watermark embedding process.
In our taxonomy, spoofing is usually downstream of stealing because the attacker must reproduce enough of the watermark signal to trigger detection.
We therefore analyze spoofing-related methods separately from direct scrubbing attacks.

\mypara{\JSVSp~\cite{jovanovic2024watermark}}
\JSVSp uses context-dependent statistics recovered from \JSVSt for detector-targeted generation.
Instead of avoiding watermarked patterns, the attack biases generation toward them.
The goal is to make unwatermarked text resemble outputs from the watermarked generator.
This creates a false-attribution risk because the detector may accept forged text as authentic.
We classify \JSVSp as spoofing because it uses recovered watermark behavior to trigger detection.

\mypara{\DEMARKSp~\cite{chen2024demark}}
\DEMARKSp uses the watermark rules recovered by \DEMARKSt during generation.
The attacker favors tokens that the hidden watermark mechanism is expected to favor.
This pushes the generated continuation toward the detector's expected watermark pattern.
Unlike \DEMARKSc, which avoids watermark-favored tokens, \DEMARKSp uses the same recovered information to imitate the watermark signal.
We classify \DEMARKSp as spoofing because it forges text that can be falsely accepted as watermarked.

\section{\sysname}
\label{sec:system_design}

\begin{figure*}[t]
    \centering
    \includegraphics[width=0.85\textwidth]{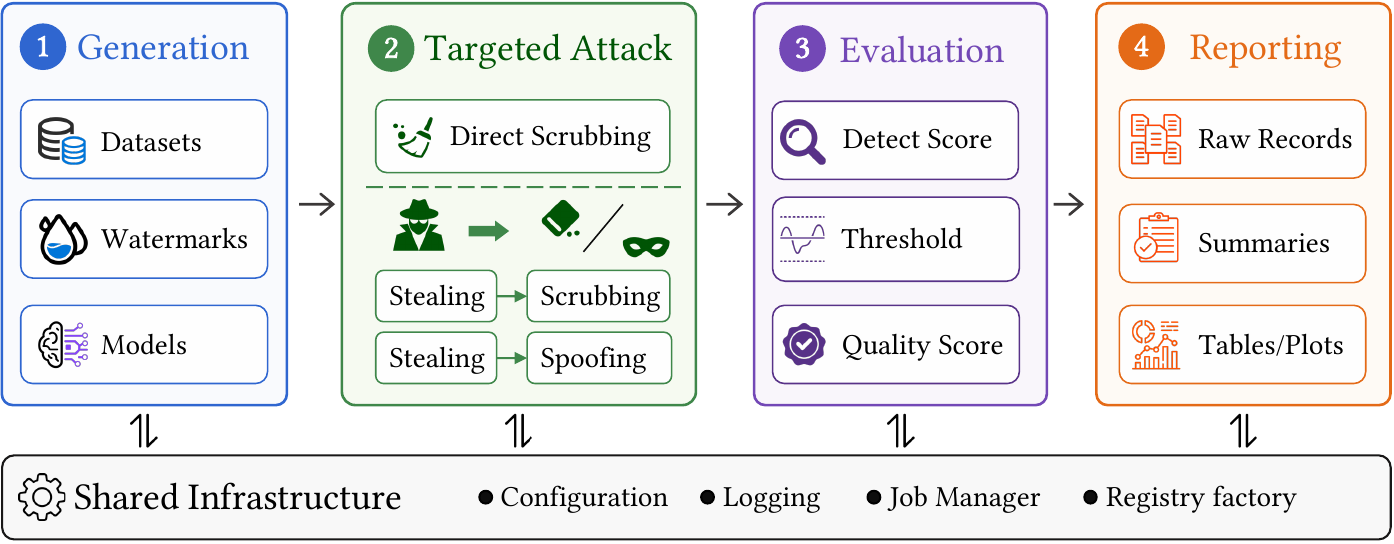}
    \Description{Pipeline overview showing how benchmark configurations select datasets, watermark methods, victim models, attacks, and evaluation metrics in the system.}
    \caption{System overview of \sysname.}
    \label{fig:system_overview}
\end{figure*}

In this section, we introduce \sysname, a modular toolkit for evaluating adversarial attacks against LLM watermarks.
\sysname operationalizes the unified framework in \autoref{sec:taxonomy} and turns it into an executable evaluation pipeline.
\sysname runs attack implementations and makes their results comparable under explicit capability assumptions.
In particular, \sysname jointly supports \textit{stealing}, \textit{scrubbing}, and \textit{spoofing}, while keeping direct scrubbing attacks separate from stronger attacks that first recover reusable watermark-related artifacts.
\autoref{fig:system_overview} presents the system overview.

Existing studies on LLM watermark evaluation mainly focus on implementing watermark methods or reporting benchmark results~\cite{pan2024markllm,tu2024waterbench,liang2025watermark}.
Attack evaluation requires more than collecting attack scripts.
Different watermark methods expose different generation interfaces, detector APIs, score directions, and threshold choices.
Different attacks also require different inputs, access assumptions, recovered artifacts, and downstream goals.
Without a shared toolkit, these differences can easily lead to results that are difficult to reproduce or compare across papers.
\sysname addresses this gap by providing a configuration-driven pipeline that loads common components, executes attacks under explicit capability assumptions, and reports attack success together with text quality.

\mypara{Modules}
\sysname consists of four main modules: input and generation, attack orchestration, detection and metrics, analysis and reporting.

\textit{Input and Generation.}
This module loads prompt datasets, target LLMs, watermark methods, and generation configurations.
It builds on existing watermark implementations and wraps them into a common generation pipeline.
For each run, it records the prompt, the watermarked text, the watermark configuration, and the decoding metadata needed by later modules.
This standardized input format decouples downstream attacks and metrics from watermark-specific implementation details.

\textit{Attack Orchestration.}
This module executes attacks according to their capability assumptions.
Direct scrubbing attacks take watermarked text as input and produce attacked text under a shared removal protocol.
Stealing-based methods first run a recovery stage, such as querying, estimation, or distillation, and then pass the recovered artifact to downstream scrubbing or spoofing.
The module records whether an attack uses the original prompt, victim-side scores, helper models, or reusable recovered artifacts.
This makes the comparison boundary explicit at execution time rather than leaving it implicit in separate attack scripts.

\textit{Detection and Metrics.}
This module applies the detector associated with each watermark method to natural, watermarked, and attacked outputs.
Because detectors may produce scores with different scales or directions, \sysname normalizes score direction when needed and supports shared threshold calibration.
The module computes detection metrics, attack success metrics, text-quality metrics, and quality-constrained attack success.
This module lets \sysname compare attack success together with text quality.

\textit{Analysis and Reporting.}
This module stores per-sample records, per-run summaries, metric details, and aggregate exports.
The raw records preserve prompts, watermarked text, attacked text, detector outputs, quality metrics, and attack-specific metadata.
The summary reports aggregate detection results, quality results, sample statistics, and the full run configuration.
\sysname also maintains a registry-backed reporting layer that supports large-scale tracking, analysis scripts, and paper-table exports.
This reporting design keeps raw experimental artifacts available while allowing higher-level benchmark views to be regenerated.

\mypara{Modular Design}
\sysname is organized around configuration-loaded wrappers for attacks, watermark generators and detectors, datasets, metrics, and experiment runners.
New attacks are integrated through a shared attack interface and registered in an attack factory.
This interface supports preparation, optional recovery or learning, editing, and saving or loading learned artifacts.
Watermark generation and detection are loaded through wrapper interfaces, while datasets and metrics are managed through registries and metric managers.
This design does not require every component to share the same internal implementation.
Instead, it provides a common boundary where heterogeneous components can be loaded, executed, and reported through the same pipeline.

\mypara{Supported Workflows}
\sysname supports three main workflows.
First, it supports unified general-scrubbing evaluation.
In this workflow, comparable removal attacks are run under shared settings and summarized in a common report.
Second, it supports capability-aware analysis of stronger attacks.
In this workflow, attacks that first recover watermark-related information are evaluated together with their recovery assumptions, recovered artifacts, and downstream use.
Third, it supports plug-and-play extension.
Researchers can add new watermark methods, attack methods, models, datasets, or metrics while reusing the same generation, detection, and reporting pipeline.

\mypara{Using \sysname}
Watermark designers can use \sysname to assess whether a watermark is vulnerable to direct scrubbing or to stronger attacks that first recover watermark-related information.
Attack researchers can use \sysname to compare new attacks against existing baselines under a unified reporting protocol.
Practitioners can use \sysname to interpret watermark scrubbing success together with output utility, so that watermark removal is not mistaken for practical attack success when the resulting text is no longer useful.
Because \sysname separates direct scrubbing from stronger stealing- and spoofing-related methods, its reports make clear which results depend on additional attacker capabilities.
This makes \sysname both the implementation vehicle of our evaluation and a reusable toolkit for future studies of LLM watermark robustness.

\section{Experimental Setup}
\label{sec:experimental_setup}
\label{sec:protocol}


\subsection{Evaluation Settings}
\label{ssec:benchmark_subjects}


\mypara{LLM Watermarks}
We conduct experiments on six representative watermark schemes, 
including three logits-based schemes, \KGW, \Unigram, and \SIR;
two posterior-based schemes, \UW and \DIP;
and one sampling-based scheme, \SynthID.
\DIP is a watermark family and is unrelated to the \DIPPER paraphraser attack.
For implementation, we build on the MarkLLM repository~\cite{pan2024markllm} where available and adapt their input-output format to our protocol.

\mypara{Evaluated SynthID Implementation}
We evaluate MarkLLM's SynthID-Text implementation in \texttt{non-distortionary} mode with the mean detector.
The detector averages keyed $g$-values over channels and eligible completion n-grams, excluding repeated contexts and positions from EOS onward.
\autoref{app:synthid-sensitivity} specifies its configuration and the cross-watermark sensitivity analysis.

\mypara{Attacks}
We include five general scrubbing attacks in the main experiments: \DIPPER, \LLMP, \RW, \SA, and \SIRA.
These attacks cover paraphrasing, LLM rewriting, quality-guided search, score-aware editing, and targeted proxy-guided rewriting.
We further include five stealing-based methods in the stronger-capability analysis: \MIP, \SCTS, \DEMARK, \JSV, and \Bfour.
Their objectives and capability assumptions are summarized in \autoref{sec:taxonomy}.
For \SCTS, we implement the attack in PyTorch~\cite{paszke2019pytorch}; for the remaining attacks, we adapt the authors' official implementations to the benchmark settings.
Detailed attack defaults are reported in~\autoref{sec:appendix_impl_details}.

\mypara{Target LLMs}
The main benchmark evaluates three widely used open-source LLM families: Llama 3.1 8B Instruct~\cite{dubey2024llama3}, Qwen2.5 7B Instruct~\cite{qwen2024qwen25}, and Mistral 7B Instruct v0.3~\cite{jiang2023mistral}.
These targets cover different tokenizer and model families while keeping the matrix computationally tractable.
All local models are loaded from fixed checkpoints to avoid inconsistencies caused by remote model updates.
For the instruction-tuning sensitivity study, we additionally compare Qwen2.5-7B Base and Instruct checkpoints under the same named \KGW settings.

\mypara{Prompt Datasets}
We conduct experiments on three prompt datasets: 
C4~\cite{raffel2020exploring}, Dolly-15K~\cite{databricks2023dolly15k}, and MMW BookReport, the book-report prompt set from Mark My Words (MMW)~\cite{piet2023mark}.
We select 100 prompts per dataset.
To better match instruction-tuned models, we prepend an instruction-style continuation prompt to each C4 sample.

\mypara{Sample Universe}
The main matrix uses 26,096 recorded outputs from 261 runs conducted in April 2026.
Each attack block is computed on its own completed-output population; execution completion includes empty or otherwise invalid outputs.
Generated source texts and detector thresholds can differ across attack runs, so cross-method comparisons aggregate configuration-level rates.
Within a run, paired detection refers to the same sample before and after attack.
The records include 35 samples with insufficient context for SIR detection and two empty SA outputs with unavailable after-detection.
Structural exclusions are marked ``N/A''; undefined measured rates are marked ``--''.

\subsection{Evaluation Metrics}
\label{ssec:metrics}


\mypara{Robustness Metrics}
True positive rate (TPR) is the fraction of watermarked texts detected as watermarked.
False positive rate (FPR) is the fraction of unwatermarked texts incorrectly detected as watermarked.
Detector calibration targets an FPR of 1\%; results use the decision and threshold recorded for each run.
This is a calibration target, not a separately measured test-set FPR.
When a detector's raw score uses the opposite direction, we invert the decision direction so that all reported results share the same interpretation.
Let $C$ contain the completed-output sample IDs of a run.
Define $S_i=1$ only when both detector states are present and sample $i$ moves from detected to undetected; otherwise it earns no removal credit.
Let $C_{+,\mathrm{pair}}$ contain samples with both states observed and a positive before-state, and let $P_i$ be the recorded Prometheus score.
We compute
\begin{align}
\ASRobs &= \frac{\sum_{i\in C}S_i}{|C_{+,\mathrm{pair}}|},\\
\QSRobs@u &= \frac{\sum_{i\in C}\indicator[S_i=1,\ P_i\ge u]}{|C|}.
\end{align}
Unavailable quality or detector evidence cannot pass the joint-success gate.
ASR uses only before-positive samples with both detector states observed; it is undefined when this set is empty.
TPR$_b$ is the positive fraction among all available before-detection results, regardless of after-detection availability.
ASR macro-averages and ranks use only cells with defined ASR; QSR macro-averages use all available cells, including those with zero detected inputs.
Each included cell receives equal weight, and figure data report the contributing cell counts.
The ASR--QSR difference therefore combines eligibility, measurement coverage, and the quality gate.

\mypara{Text-Quality Metrics}
We use Prometheus (Prom)~\cite{kim2024prometheus2} to assess output quality at thresholds 2, 3, and 4.
BERTScore with \texttt{roberta-large} and PSP provide complementary similarity measures.
QSR denotes removal jointly satisfying the specified automatic quality gate; QSR@3 is the default.
BERTScore and PSP compare full-text fields that include a shared prompt when present (\autoref{app:general_scrubbing_matrices}).
Quality evaluation uses automatic scores; supplementary content screening uses deterministic rules and Codex-assisted source--output inspection, with no independent human annotation.
Source adequacy and preservation are assessed separately: a faithful edit can retain an inadequate source.

\subsection{Evaluation Matrix And Reporting Scope}
\label{ssec:evaluation_matrix}

Each benchmark run is defined by a tuple $(D, M, W, A)$, where $D$ represents a prompt dataset, $M$ a target LLM, $W$ a watermark scheme, and $A$ an attack method.
To ensure readability and protocol control, the main text focuses on the canonical C4 slice.
Additional Dolly-15K and MMW BookReport slices are included in the appendix to examine dataset dependence.

We report general scrubbing attacks as the main comparison surface because they share the same downstream objective: given watermarked text, produce an attacked text with weakened watermark evidence while preserving utility.
Stealing- and spoofing-related methods are reported separately because they require additional capabilities, such as query access, query budgets, prebuilt corpora, recovered artifacts, or attack-side generation control.
This shared-protocol comparison answers how robust watermark families are under general scrubbing.
The stronger-capability analysis asks when recovered watermark-related information changes the risk picture through targeted scrubbing or spoofing.

\section{Experimental Results}
\label{sec:experimental_results}

In this section, we answer five empirical questions.
\begin{enumerate}[label=\textbf{RQ\arabic*.}, leftmargin=*, itemsep=0.2em]
    \item How robust are LLM watermarks under the shared general-scrubbing protocol?
    \item Are general scrubbing attacks broad quality-aware baselines or watermark-family specialists?
    \item How does Prometheus-based utility filtering change attack rankings?
    \item How do the Base--Instruct observations vary with prompting protocol?
    \item How does recovered watermark-related information change scrubbing and spoofing risk?
\end{enumerate}
The first three questions use the shared general-scrubbing benchmark, where each attack receives watermarked text and outputs attacked text under the same reporting protocol.
RQ1 measures robustness under this shared protocol.
RQ2 separates broad quality-aware baselines from family-aligned specialists.
RQ3 asks whether rankings survive Prometheus-gated utility filtering.
RQ4 compares Base--Instruct outcomes under two prompting protocols.
RQ5 studies whether recovered watermark-related information changes downstream scrubbing and spoofing risk.
We report RQ5 separately from the general-scrubbing leaderboard because its access assumptions are different.

\begin{table*}[t]
\centering
\small
\caption{C4 general-scrubbing results across watermark families, victim models, and attacks.}
\label{tab:rq1-general-scrub-detect}
\setlength{\tabcolsep}{3.5pt}
\renewcommand{\arraystretch}{0.85}

\begin{tabular}{ll|ccc|ccc|ccc|ccc|ccc}
\toprule
\multirow{2}{*}{Watermark} & \multirow{2}{*}{Model} & \multicolumn{3}{c|}{DIPPER} & \multicolumn{3}{c|}{LLMP} & \multicolumn{3}{c|}{SIRA} & \multicolumn{3}{c|}{RW} & \multicolumn{3}{c}{SA} \\
\cmidrule(lr){3-5}\cmidrule(lr){6-8}\cmidrule(lr){9-11}\cmidrule(lr){12-14}\cmidrule(lr){15-17}
& & TPR$_b\uparrow$ & ASR$\uparrow$ & QSR$\uparrow$ & TPR$_b\uparrow$ & ASR$\uparrow$ & QSR$\uparrow$ & TPR$_b\uparrow$ & ASR$\uparrow$ & QSR$\uparrow$ & TPR$_b\uparrow$ & ASR$\uparrow$ & QSR$\uparrow$ & TPR$_b\uparrow$ & ASR$\uparrow$ & QSR$\uparrow$ \\
\midrule
\multirow{3}{*}{\KGW} & Llama & 0.93 & 0.78 & \textbf{0.63} & 0.92 & 0.67 & 0.57 & 0.93 & 0.82 & 0.42 & 0.96 & 0.61 & 0.35 & 1.00 & 0.46 & 0.26 \\
 & Qwen & 0.96 & 0.75 & 0.61 & 1.00 & 0.88 & 0.72 & 0.96 & 0.91 & \textbf{0.74} & 0.99 & 0.62 & 0.26 & 0.95 & 0.68 & 0.33 \\
 & Mistral & 0.94 & 0.88 & 0.71 & 0.97 & 0.82 & \textbf{0.78} & 0.94 & 0.89 & 0.44 & 0.95 & 0.81 & 0.39 & N/A & N/A & N/A \\
\midrule
\multirow{3}{*}{\Unigram} & Llama & 0.72 & 0.75 & 0.46 & 0.77 & 0.75 & \textbf{0.53} & 0.72 & 0.96 & 0.43 & 0.73 & 0.88 & 0.30 & 0.73 & 0.49 & 0.19 \\
 & Qwen & 0.81 & 0.85 & 0.57 & 0.86 & 0.98 & \textbf{0.78} & 0.81 & 0.95 & 0.65 & 0.80 & 0.64 & 0.24 & 0.75 & 0.89 & 0.22 \\
 & Mistral & 0.90 & 0.60 & 0.50 & 0.81 & 0.78 & \textbf{0.58} & 0.90 & 0.88 & 0.47 & 0.91 & 0.59 & 0.31 & N/A & N/A & N/A \\
\midrule
\multirow{3}{*}{\SIR} & Llama & 0.62 & 0.62 & 0.33 & 0.55 & 0.78 & \textbf{0.40} & 0.51 & 0.84 & 0.27 & 0.49 & 0.71 & 0.23 & 0.59 & 0.76 & 0.26 \\
 & Qwen & 0.69 & 0.79 & 0.41 & 0.70 & 0.96 & \textbf{0.63} & 0.77 & 0.95 & 0.62 & 0.71 & 0.76 & 0.32 & 0.70 & 0.91 & 0.24 \\
 & Mistral & 0.91 & 0.55 & 0.43 & 0.91 & 0.69 & \textbf{0.61} & 0.91 & 0.73 & 0.38 & 0.93 & 0.80 & 0.32 & N/A & N/A & N/A \\
\midrule
\multirow{3}{*}{\UW} & Llama & 0.90 & 0.93 & 0.71 & 0.85 & 0.92 & \textbf{0.75} & 0.90 & 0.96 & 0.47 & 0.87 & 0.78 & 0.48 & 0.90 & 0.73 & 0.24 \\
 & Qwen & 0.84 & 0.99 & 0.68 & 0.83 & 1.00 & \textbf{0.76} & 0.84 & 1.00 & 0.71 & 0.75 & 0.93 & 0.37 & 0.82 & 1.00 & 0.26 \\
 & Mistral & 0.95 & 0.96 & \textbf{0.84} & 0.94 & 0.93 & 0.82 & 0.95 & 0.96 & 0.43 & 0.98 & 0.84 & 0.50 & 0.95 & 0.99 & 0.46 \\
\midrule
\multirow{3}{*}{\DIP} & Llama & 0.57 & 1.00 & 0.49 & 0.72 & 0.90 & \textbf{0.61} & 0.57 & 0.93 & 0.37 & 0.70 & 0.89 & 0.40 & 0.75 & 0.97 & 0.19 \\
 & Qwen & 0.82 & 0.94 & 0.59 & 0.81 & 0.96 & \textbf{0.72} & 0.82 & 0.96 & 0.69 & 0.78 & 0.88 & 0.34 & 0.86 & 0.99 & 0.32 \\
 & Mistral & 0.85 & 0.94 & 0.68 & 0.84 & 0.95 & \textbf{0.76} & 0.85 & 0.92 & 0.35 & 0.85 & 0.85 & 0.37 & 0.78 & 0.99 & 0.46 \\
\midrule
\multirow{3}{*}{\SynthID} & Llama & 0.95 & 0.80 & 0.67 & 0.99 & 0.89 & \textbf{0.85} & 0.95 & 0.87 & 0.43 & 0.96 & 0.86 & 0.47 & 0.96 & 0.95 & 0.22 \\
 & Qwen & 0.98 & 0.85 & 0.70 & 0.93 & 0.97 & 0.85 & 0.98 & 0.98 & \textbf{0.87} & 0.94 & 0.88 & 0.42 & 0.95 & 1.00 & 0.27 \\
 & Mistral & 0.97 & 0.69 & 0.58 & 0.99 & 0.81 & \textbf{0.76} & 0.97 & 0.74 & 0.40 & 0.96 & 0.82 & 0.42 & 0.97 & 0.91 & 0.38 \\
\bottomrule
\end{tabular}
\end{table*}

\subsection{RQ1: Overall General-Scrubbing Robustness}

We first evaluate general scrubbing on the canonical C4 slice.
\autoref{tab:rq1-general-scrub-detect} reports pre-attack detectability, raw attack success, and quality-constrained success for each watermark--victim--attack combination.
We abbreviate pre-attack TPR as TPR$_b$.
Higher ASR and QSR mean stronger attacks.
Bold marks the row-wise best QSR.
We use ``N/A'' for combinations excluded by the benchmark protocol.
In this table, ``N/A'' appears for \SA under selected Mistral settings because \SA requires a tokenizer-aligned helper model, and our Mistral setup does not include a matching small helper model.
The relevant distinctions are baseline detectability, paired detector evasion, and automatic-quality-gated success.

Two patterns stand out.
First, attack success has to be interpreted relative to the pre-attack detector strength.
Some watermark families start from much stronger detectability than others, so a high ASR on a weakly detectable setting is not equivalent to breaking a strongly embedded watermark.
Second, no general scrubbing attack dominates all axes.
\SIRA leads the recorded C4 removal rate, while \LLMP leads joint success under the automatic Prometheus gate.
\DIPPER remains the closest quality-aware alternative on C4.

\begin{figure*}[t]
    \centering
    \includegraphics[width=\textwidth]{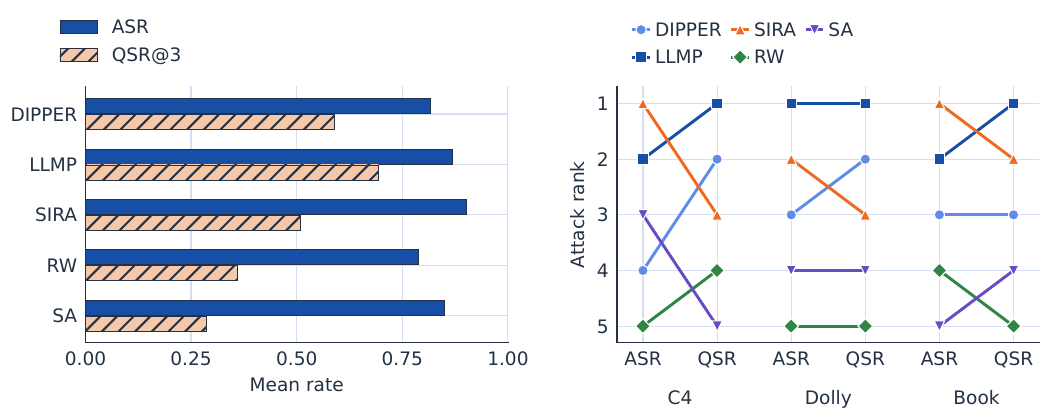}
    \Description{Two-panel figure comparing mean ASR and QSR@3 on C4 and their corresponding attack rankings across dataset slices.}
    \caption{Mean ASR and QSR@3 on C4 and attack ranks across datasets. ASR averages exclude undefined cells; QSR averages include all available cells.}
    \label{fig:rq1-asr-qsr-gap}
\end{figure*}

\autoref{fig:rq1-asr-qsr-gap} visualizes the difference between the two reported metrics.
On C4, \SA illustrates the danger of using ASR alone.
It often drives detector scores down, but fewer records receive joint credit under the automatic gate and joint-rate denominator.
\LLMP and \DIPPER lose less of their apparent success when QSR is applied.
Across dataset slices, attack rankings differ between ASR and QSR@3.

\takeaway{On C4, \SIRA achieves the highest mean ASR, whereas \LLMP achieves the highest mean QSR@3.}

\subsection{RQ2: Family-Specific Robustness Patterns}
RQ2 asks whether general scrubbing attacks are broadly effective across watermark families or mainly strong when their mechanism aligns with a specific watermark.
This question matters because overall averages can hide two different behaviors: a stable broad baseline and a narrow family specialist.

\begin{figure}[tbp]
    \centering
    \includegraphics[width=\linewidth]{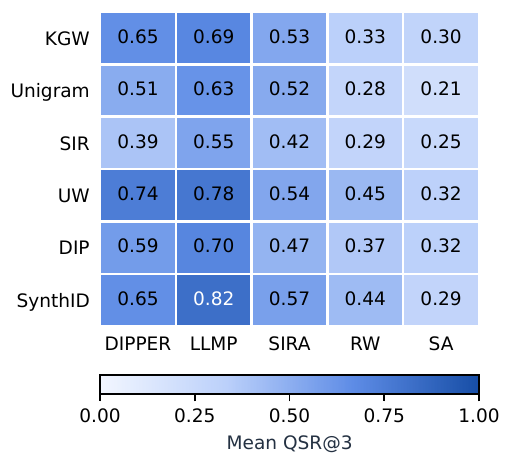}
    \Description{Heatmap comparing mean QSR@3 across general scrubbing attacks and watermark families on the C4 slice.}
    \caption{Mean QSR@3 by watermark on C4, averaging available victim-model cells. UW denotes Unbiased.}
    \label{fig:rq2-family-qsr-heatmap}
\end{figure}

To answer RQ2, we read the C4 matrix by watermark family and summarize the interaction in~\autoref{fig:rq2-family-qsr-heatmap}.
The main result is that quality-aware attack rankings are less fragmented than raw removal suggests.
\LLMP has the highest mean QSR@3 for every C4 watermark family.
The family-conditioned view still matters because the margin changes by family.
\UW and \KGW leave \DIPPER close to \LLMP, while \SynthID shows the widest \LLMP lead.
\Unigram, \SIR, and \DIP fall between these two cases.

\begin{figure}[tbp]
    \centering
    \includegraphics[width=\linewidth]{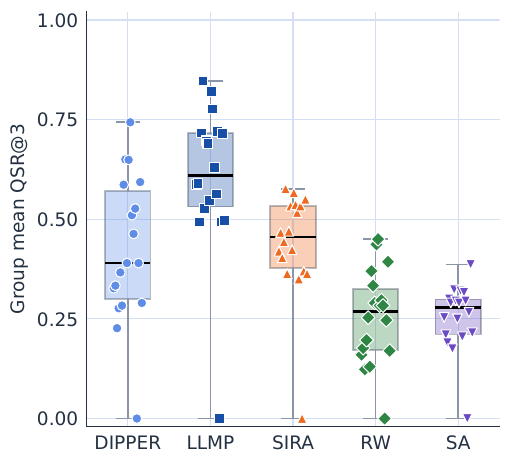}
    \Description{Box plot with overlaid points showing each attack's QSR@3 distribution over dataset--watermark cells.}
    \caption{QSR@3 over 18 dataset--watermark groups per attack, each averaged over available models. Points show every group; boxes indicate quartiles and whiskers span the observed range.}
    \label{fig:rq2-generality-distribution}
\end{figure}

\autoref{fig:rq2-generality-distribution} adds a supporting view over dataset--watermark cells.
Unlike~\autoref{fig:rq2-family-qsr-heatmap}, which summarizes family-level averages, this figure shows how stable each attack is across individual $(D, W)$ conditions.
Each point is one cell, and each box summarizes the QSR@3 distribution for one attack.
\LLMP has the highest quality-aware success distribution across these cells.
Other attacks either sit lower overall or show wider variation.

This interaction view changes how we should interpret attack rankings.
Overall averages identify \LLMP as the strongest quality-aware attack on C4.
Family-conditioned analysis explains where that lead is narrow and where it is robust.
\SIRA leads mean ASR on C4, while \LLMP leads mean QSR@3 in every C4 watermark family.
Family conditioning mainly changes the margin, not the top QSR attack on C4.

\takeaway{\LLMP is the strongest quality-aware general-scrubbing baseline on C4, with the narrowest margins under \UW and \KGW.}

\mypara{SynthID: Shared Trend and Model-Level Exceptions}
\SynthID shares the C4 family-average trend, but its Qwen cell gives \SIRA $\QSRobs@3=0.87$ versus 0.85 for \LLMP; the Llama values are 0.43 and 0.85.
Across these two models, \SA removes 186 of 191 recorded pre-positive passages, but only 49 removals pass Prometheus 3.
For \LLMP, 178 of 192 recorded pre-positive passages are removed and 170 pass the gate.
For SynthID--\LLMP, C4 has 192/200 pre-positive records, 178 removals, and 170 gated successes, whereas Dolly has 113/200, 112, and 104.
Dolly's lower joint rate, 0.52 versus 0.85, accompanies near-complete removal of its smaller eligible set.

\subsection{RQ3: Utility-Aware Robustness Trade-Offs}

RQ3 asks how the recorded joint-success rates change as the automatic acceptance threshold is tightened.

For readability, the main text uses two C4 views: threshold sensitivity and attack-conditioned cell variation.
\autoref{app:general_scrubbing_matrices} reports the full per-dataset matrices for detection, QSR, BERTScore, PSP, and Prometheus.

\begin{figure}[t]
    \centering
    \includegraphics[width=\linewidth]{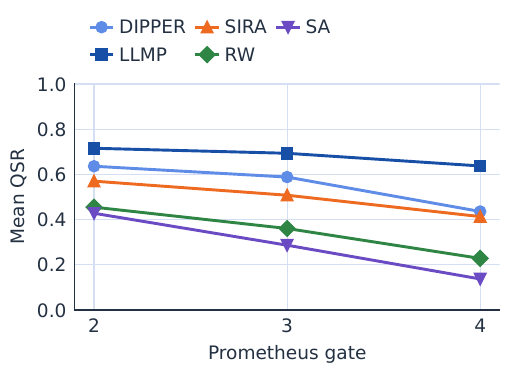}
    \Description{Mean QSR curves for five attacks as the Prometheus quality threshold increases from 2 to 4.}
    \caption{Mean QSR on C4 at Prometheus gates 2, 3, and 4. Each method averages its available victim-model cells.}
    \label{fig:rq3-threshold-sensitivity}
\end{figure}

\autoref{fig:rq3-threshold-sensitivity} makes the threshold effect explicit.
\LLMP retains the largest fraction of its QSR under stricter quality thresholds.
\LLMP also remains the strongest attack at each reported threshold.
\SA degrades much faster.
\SIRA has the strongest raw ASR, but its quality-aware success drops once the utility gate tightens.
\DIPPER remains the closest quality-aware alternative to \LLMP on C4.
The counterintuitive point is that a stronger detector-side attack can become weaker under a stricter utility gate.

\begin{figure*}[t]
    \centering
    \includegraphics[width=\textwidth]{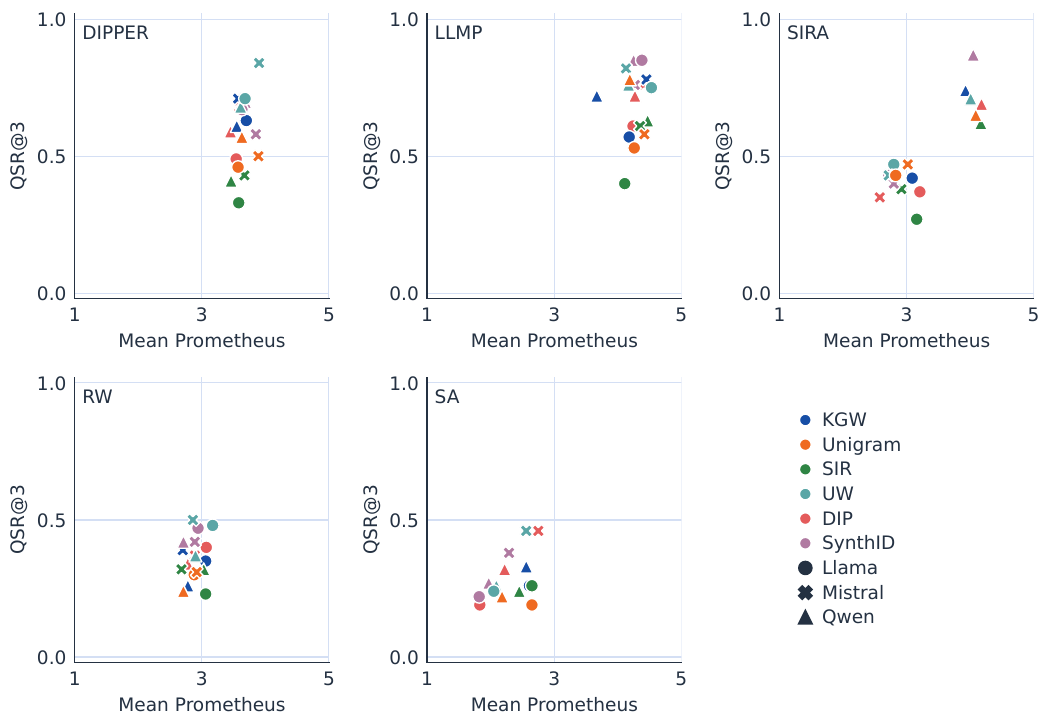}
    \Description{Small-multiple scatter plots showing Prometheus and QSR@3 on C4. Colors identify watermark families; circle, cross, and triangle markers identify Llama, Mistral, and Qwen respectively.}
    \caption{Prometheus scores and QSR@3 on C4. Each point is one victim-model--watermark cell. Color identifies the watermark; shape identifies the model.}
    \label{fig:rq3-attack-conditioned-tradeoff}
\end{figure*}

\autoref{fig:rq3-attack-conditioned-tradeoff} adds a cell-level view.
Each point is one watermark--victim-model unit.
The plot shows how automatic quality scores and joint success vary across watermark and victim-model conditions.
It complements the threshold view by exposing the variation hidden by aggregate attack rankings.

\takeaway{\LLMP achieves the highest mean QSR on C4 at all three Prometheus gates.}

\mypara{Cross-Watermark Sensitivity on Common Support}
The supplementary analysis fixes Prometheus gates 2, 3, and 4 and uses 180 cells with the same two-model support for all six watermarks and five attacks, keeping the three datasets separate.
For SynthID on C4, \LLMP's joint rates are 0.875, 0.850, and 0.770; its margins over the best other attack are 12.11, 16.65, and 24.50 percentage points.
\LLMP leads each C4 watermark family at all three gates, and each family's margin is larger at 4 than at 2.
SynthID has the largest C4 margin at gate 4.
The Qwen exception favors \SIRA at gates 2 and 3 but \LLMP at gate 4, 0.78 versus 0.76.
\autoref{tab:synthid-gate-sensitivity} reports the corresponding results for all six watermark families.

\subsection{RQ4: Instruction-Tuning Sensitivity}
\label{ssec:instruction_tuning_ablation}

RQ4 asks how the Base--Instruct difference in watermark robustness varies with prompting protocol.
We compare \texttt{Qwen2.5-7B} and \texttt{Qwen2.5-7B-Instruct} on C4 at two \KGW settings, $h{=}1,\gamma{=}0.25$ and $h{=}1,\gamma{=}0.5$, using \LLMP and \SIRA.
The \emph{native} condition uses automatic template selection; the \emph{wrapper-disabled} condition omits the chat wrapper.
The comparison contains 16 checkpoint--prompting--watermark--attack combinations.
Each run uses its generated source texts and recorded detector threshold, so the table reports configuration-level Instruct-minus-Base differences.

\begin{table*}[t]
\centering
\small
\caption{Instruct-minus-Base differences on Qwen2.5-7B/KGW under native and wrapper-disabled prompting.}
\label{tab:instruction-tuning-ablation}
\setlength{\tabcolsep}{8pt}
\renewcommand{\arraystretch}{0.85}
\begin{tabular}{ll c cc cc}
\toprule
Setting & Protocol & $\Delta$TPR$_b$ & $\Delta$ASR$_{\text{LLMP}}$ & $\Delta$ASR$_{\text{SIRA}}$ & $\Delta$QSR$_{\text{LLMP}}$ & $\Delta$QSR$_{\text{SIRA}}$ \\
\midrule
$h{=}1,\gamma{=}0.25$ & Native & -0.045 & +0.106 & +0.046 & +0.170 & +0.160 \\
$h{=}1,\gamma{=}0.25$ & Wrapper-disabled & +0.000 & -0.070 & +0.070 & +0.020 & +0.090 \\
$h{=}1,\gamma{=}0.5$ & Native & -0.015 & +0.061 & +0.059 & +0.180 & +0.210 \\
$h{=}1,\gamma{=}0.5$ & Wrapper-disabled & +0.020 & -0.028 & -0.017 & +0.060 & +0.120 \\
\bottomrule
\end{tabular}
\end{table*}

Under native prompting, the Instruct-minus-Base pre-detection differences are $-0.045$ for $\gamma=0.25$ and $-0.015$ for $\gamma=0.5$.
For $\gamma=0.25$, LLMP and SIRA have native QSR differences of $+0.170$ and $+0.160$; disabling the wrapper reduces these descriptive differences to $+0.020$ and $+0.090$.
For $\gamma=0.5$, the native QSR differences are $+0.180$ and $+0.210$, compared with $+0.060$ and $+0.120$ without the wrapper.
The corresponding ASR differences change sign in three of four wrapper-disabled entries.

\takeaway{In the evaluated Qwen2.5-7B/\KGW runs, the Instruct--Base QSR gap is smaller under wrapper-disabled prompting than under native prompting.}

\begin{table*}[t]
\centering\small
\caption{Stealing-based scrubbing on C4/Llama under four KGW settings. General comparators are the highest QSR@3 values among general scrubbers at each setting.}
\label{tab:c4-steal-scrub}
\setlength{\tabcolsep}{5pt}
\begin{tabular}{llrrrrrr}
\toprule
KGW & Method & $n$ & ASR & QSR@3 & Best general QSR & $\Delta$QSR & Prom \\
\midrule
$h=1,\gamma=0.25$ & \Bfour & 100 & 0.78 & 0.72 & \DIPPER 0.59 & +0.13 & 4.35 \\
 & \MIP & 100 & 0.17 & 0.10 & \DIPPER 0.59 & -0.49 & 3.21 \\
 & \DEMARK & 100 & 0.61 & 0.42 & \DIPPER 0.59 & -0.17 & 2.91 \\
 & \SCTS & 100 & 0.22 & 0.20 & \DIPPER 0.59 & -0.39 & 4.11 \\
 & \JSV & 100 & 0.75 & 0.42 & \DIPPER 0.59 & -0.17 & 3.01 \\
\addlinespace[2pt]
$h=1,\gamma=0.5$ & \Bfour & 100 & 0.71 & 0.60 & \DIPPER 0.63 & -0.03 & 4.32 \\
 & \MIP & 100 & 0.33 & 0.23 & \DIPPER 0.63 & -0.40 & 3.04 \\
 & \DEMARK & 100 & 0.38 & 0.27 & \DIPPER 0.63 & -0.36 & 2.97 \\
 & \SCTS & 100 & 0.26 & 0.23 & \DIPPER 0.63 & -0.40 & 4.38 \\
 & \JSV & 99 & 0.84 & 0.41 & \DIPPER 0.63 & -0.22 & 2.89 \\
\addlinespace[2pt]
$h=3,\gamma=0.25$ & \Bfour & 100 & 0.89 & 0.78 & \DIPPER 0.80 & -0.02 & 4.22 \\
 & \MIP & 100 & 0.72 & 0.44 & \DIPPER 0.80 & -0.36 & 3.16 \\
 & \DEMARK & 100 & 0.00 & 0.00 & \DIPPER 0.80 & -0.80 & 2.79 \\
 & \SCTS & 100 & 0.49 & 0.46 & \DIPPER 0.80 & -0.34 & 4.46 \\
 & \JSV & 100 & 0.88 & 0.53 & \DIPPER 0.80 & -0.27 & 3.05 \\
\addlinespace[2pt]
$h=3,\gamma=0.5$ & \Bfour & 100 & 0.89 & 0.83 & \DIPPER 0.80 & +0.03 & 4.31 \\
 & \MIP & 100 & 0.78 & 0.55 & \DIPPER 0.80 & -0.25 & 3.09 \\
 & \DEMARK & 100 & 0.00 & 0.00 & \DIPPER 0.80 & -0.80 & 2.64 \\
 & \SCTS & 100 & 0.39 & 0.37 & \DIPPER 0.80 & -0.43 & 4.48 \\
 & \JSV & 99 & 0.83 & 0.48 & \DIPPER 0.80 & -0.32 & 3.13 \\
\addlinespace[2pt]
\bottomrule
\end{tabular}
\end{table*}


\subsection{RQ5: Stealing-Based Attacks}

RQ5 asks whether recovered watermark information changes the risk picture beyond the general-scrubbing benchmark.
Unlike RQ1--RQ3, these attacks assume that the attacker first steals or infers reusable watermark information, such as green-list structure, token preferences, or decision statistics.
We compare them with strong general-scrubbing baselines, but we do not merge them into the same leaderboard.

The five stealing-based methods share \KGW as a target, enabling a comparison across four settings with $h\in\{1,3\}$, $\gamma\in\{0.25,0.5\}$, and $\delta=2$.
Mechanistically, $h$ defines how much prior context enters the favored-token rule, while $\gamma$ defines the green-token base rate.
Recovered artifacts guide scrubbing by avoiding or replacing watermark-favored tokens; spoofing-facing use would reverse this direction and favor such tokens during generation.
The runs use method-specific helper models, source texts, and recovery budgets.
Execution time and recovery resources are reported separately in \autoref{tab:appendix_operational_profile}.

\autoref{tab:c4-steal-scrub} compares representative stealing-based scrubbing methods against the best general-scrubbing baselines on C4.
It reports paired-removal ASR, QSR@3, sample counts, Prometheus scores, and differences from the highest general-scrubbing QSR at each setting.

\Bfour achieves the highest QSR@3 among the stealing-based scrubbers in all four settings, with rates from 0.60 to 0.83.
Its QSR@3 exceeds the best general-scrubbing comparator by 0.13 and 0.03 at two settings and is lower by 0.03 and 0.02 at the other two.

The ASR and QSR comparisons show different outcomes for stealing-based attacks.
\JSV often approaches the best general-scrubbing baseline in ASR, while its QSR@3 remains lower.
\SCTS shows the opposite failure mode: it preserves high Prometheus quality, yet its QSR remains below the strongest general-scrubbing baseline because removal is too weak.
\MIP and \DEMARK are more setting-sensitive and do not provide a consistent gain across the four \KGW settings.

Spoofing is the separate case where recovered information is used for false attribution rather than removal.
We run a small \KGW spoofing feasibility check for \JSVSp and \DEMARKSp, reported in \autoref{tab:spoof-feasibility-kgw}.
$\mathrm{QSSR}@3$ counts forged outputs that trigger the watermark detector and receive Prometheus $\ge 3$ against their paired watermarked continuation.
The eight spoofing runs each contain 100 outputs; natural-generation detector acceptance is zero in each run.
\DEMARKSp reaches 98--100\% detector acceptance and 47--64\% QSSR@3 across the four settings.
\JSVSp reaches 0--1\% detector acceptance and zero QSSR@3.

\takeaway{\Bfour leads stealing-based QSR@3 in all four \KGW settings; \DEMARKSp attains 98--100\% detector acceptance and 47--64\% QSSR@3.}

\subsection{Supplementary Output Screening}
\label{ssec:evidence_audit}
We supplement the automatic metrics with output screening against the task and source text.
Deterministic rules identify empty and punctuation-only outputs; length reduction and repetition flag texts for Codex-assisted inspection.
Across 26,096 outputs, we inspect 251 source--output pairs and identify 254 invalid outputs, including 19 deterministic failures.
An invalid output contributes zero to screened QSR while remaining in its original denominator.
This excludes 20 QSR@3 successes across 20 cells, and \LLMP retains the highest dataset-level mean QSR@3 on C4, Dolly-15K, and BookReport.
\autoref{app:content_validity} gives the screening criteria, review coverage, and per-method changes.

\subsection{Scope And Limitations}
\label{ssec:scope_limitations}

The general-scrubbing matrix covers six watermark families, three instruction-tuned model families, and three prompt datasets.
The Base--Instruct and recovered-information studies cover \KGW under their specified prompting and access conditions.
These configuration-level comparisons use the run-specific sources, detector thresholds, and resource scopes defined in \autoref{ssec:metrics} and \autoref{sec:appendix_impl_details}.

\mypara{Interpretation and Decision Value}
A low joint-success rate can reflect few detectable inputs, low removal among eligible inputs, a strict automatic gate, or missing measurements.
These cases lead to different security decisions and should not be collapsed into a watermark ranking.
Evaluators should inspect baseline eligibility and paired transitions before choosing attack tests, and verify source identity before pairing methods for uncertainty estimation.

\section{Related Work}
\label{sec:related_work}

Recent LLM watermark benchmarks and toolkits have improved the reproducibility of watermark implementation and evaluation.
MarkLLM is a representative implementation-oriented toolkit, providing unified interfaces for implementing watermark methods and evaluating generation, detection, quality, and robustness behavior~\cite{pan2024markllm}.
Complementing this toolkit perspective, WaterBench studies fair comparison of watermark methods under aligned watermarking strength, with particular attention to detection--generation trade-offs and instruction-following quality~\cite{tu2024waterbench}.
From a robustness perspective, WaterPark is the closest prior platform: it integrates a broad suite of removal attacks to evaluate the resilience of LLM watermarkers under adversarial perturbations~\cite{liang2025watermark}.
Other work broadens the evaluation criteria.
CEFW proposes a multi-dimensional scoring framework that covers detectability, text quality, usability, robustness, and imperceptibility~\cite{zhang2025cefw}.
MarkMyWords further benchmarks watermark schemes across quality, size, and tamper-resistance under practical attacks~\cite{piet2023mark}.

These frameworks are complementary to \sysname, but they are mostly organized around watermark methods rather than attack methods.
In these systems, attacks typically serve as robustness tests, quality--robustness trade-off factors, or components of a composite watermark score.
WaterPark is the closest exception because it substantially expands the robustness evaluation.
WaterPark also examines semantic preservation and watermark design factors.
Its main evaluation target remains the resilience of watermarkers under attacks, not a capability-stratified benchmark of attack methods themselves.
Prior frameworks also provide limited support for organizing stealing, spoofing, recovered artifacts, and general scrubbing as separate evaluation tracks.

\sysname addresses this attack-centric gap by asking how attack methods behave under explicit capability assumptions.
It treats attack methods as a first-class benchmark dimension.
It separates the general-scrubbing comparison from stronger stealing and spoofing analyses.
It also reports attack effectiveness together with text quality.
This makes \sysname complementary to watermark implementation and watermark-scoring frameworks: it stress-tests LLM watermarks through modern attack methods and reports the assumptions behind those attacks.

\section{Conclusion}
\label{sec:conclusion}

We presented \sysname, a general framework for evaluating adversarial attacks against LLM watermarks under a unified reporting protocol. 
The framework is built around an explicit separation between comparable general scrubbing attacks and stronger stealing-based attacks that require additional access, recovered artifacts, or attack-side control.
This design allows shared-protocol robustness comparisons while still exposing escalation risks that would be hidden by a text-only evaluation.

Across the evaluated matrix, three findings stand out. 
First, there is no universal winner across all evaluation axes. 
Second, family-specific behavior matters: overall averages are useful, but they can hide whether an attack is a broad baseline or a narrow specialist. 
Third, \SIRA leads mean removal on C4, while \LLMP leads mean Prometheus-gated joint success across watermark families and retains that lead under stricter gates.
Supplementary output screening preserves \LLMP's lead in dataset-level mean QSR@3.
The Base--Instruct comparison shows smaller QSR gaps with the prompt wrapper disabled.
Under stronger recovered-information access, \Bfour achieves the highest QSR@3 among the evaluated stealing-based scrubbers in all four \KGW settings, and \DEMARKSp achieves high detector acceptance with lower quality-gated spoofing success.

More broadly, our results argue for capability-aware watermark evaluation. 
Stronger attacks should not be ignored, but they should be reported together with the assumptions, query budgets, runtime costs, and artifact provenance that make them possible. 
We hope \sysname helps move LLM watermark evaluation toward more reproducible, interpretable, and practically meaningful security assessment.

\bibliographystyle{ACM-Reference-Format}
\bibliography{wm_refs}

\appendix
\section{Open Science}
\label{sec:appendix_open_science}

\mypara{Reproducibility Scope}
The reproducibility materials distinguish the manuscript source from the benchmark code, experimental records, and third-party assets.

We view reproducibility as a core requirement for benchmark papers.
The main contribution of \sysname is a reusable evaluation protocol that others can inspect, rerun, and extend.
To support that goal, the project is organized around explicit configuration files, versioned benchmark defaults, and shared interfaces for watermark generation, attack execution, detection, quality evaluation, and result persistence.

\mypara{Artifact Scope}
The benchmark repository contains the framework code needed to instantiate the reported evaluation pipeline, including attack adapters, watermark wrappers, detector calibration logic, experiment orchestration, quality-analysis modules, and paper-facing plotting/reporting scripts.
The configuration-driven design makes benchmark assumptions auditable: attack settings, watermark parameters, datasets, victim models, and reporting thresholds are defined in named configs rather than buried in ad hoc scripts.

\mypara{Reproducibility Assets}
The evidence manifest identifies TeX and figure dependencies, per-sample records, configurations, and analysis outputs by cryptographic hashes.
The experiment records identify the April 2026 runs, September supplementary scoring, and output-screening decisions as separate measurements.
The source archive accompanying the PDF is a manuscript archive, not a claim that all third-party models, datasets, or recovered artifacts are redistributable.
This is especially important for stealing-based attacks, where recovered artifacts and corpus provenance materially affect interpretation.

\mypara{Dependencies And Non-Redistributable Components}
Some components of the benchmark depend on external model checkpoints, third-party repositories, or datasets that may have their own licenses or access controls.
Redistribution of these components is governed by their respective licenses and access conditions; users may need to obtain third-party assets separately under their original terms.
The framework is designed to resolve local model paths through repo configuration so that experiments can be rerun on machines with different storage layouts without changing benchmark logic.

\mypara{Planned Benchmark Release}
We intend to release the framework source, configuration files, experiment manifests, and figure-generation scripts needed to reproduce the reported benchmark protocol and extend it with new attacks, watermark families, datasets, and models.
This planned release does not include third-party model checkpoints, datasets, or recovered artifacts unless their redistribution terms permit it.

\section{Ethical Considerations}
\label{sec:appendix_ethical_considerations}

This paper studies attacks on LLM watermarking systems, which creates an unavoidable dual-use tension.
Stronger attack benchmarks are necessary for understanding whether watermarking methods provide meaningful robustness in practice.
The same evaluation infrastructure can also lower the barrier to studying or operationalizing watermark removal, stealing, or spoofing strategies.
We therefore frame \sysname as a stress-testing benchmark for defensive evaluation, not as a claim that watermark removal should be optimized without qualification.

\mypara{Why This Evaluation Is Necessary}
Watermarks are often discussed as mechanisms for provenance, platform governance, and misuse mitigation.
If their robustness is evaluated only against weak or incomparable baselines, practitioners may overestimate the security they provide.
A realistic attack benchmark is therefore important for preventing false confidence, identifying failure modes before deployment, and making defender-side trade-offs visible.

\mypara{Dual-Use Risk And Mitigation}
The highest-risk part of this space is capability-escalated attacks that recover reusable watermark artifacts or support downstream spoofing.
For that reason, the paper does not collapse all attacks into a single leaderboard.
It reports stronger stealing-based threats separately and emphasizes the assumptions that make them possible, including access patterns, query budgets, runtime costs, and watermark-family alignment.
This reporting choice is itself a mitigation: it keeps the analysis focused on security interpretation rather than on decontextualized ``best attack'' claims.

\mypara{Release And Use Considerations}
Any public release of the benchmark should prioritize reproducibility for defenders and evaluators while avoiding unnecessary concentration of turnkey high-risk assets.
In practice, this means documenting assumptions, preserving artifact provenance, and distinguishing benchmark infrastructure from third-party models or recovered attack-side artifacts that may require additional controls or independent reconstruction.
Users of the benchmark should interpret the reported attacks as stress tests for robustness claims and should avoid deploying them against systems they do not own or have permission to evaluate.

\mypara{Broader Impact}
Our goal is to improve the reliability of claims about LLM watermark robustness.
We believe transparent evaluation with explicit capability assumptions is more responsible than relying on optimistic or incomparable robustness reports, especially for techniques that may inform provenance or policy decisions.
At the same time, benchmark results should not be read as evidence that watermarking is either universally sufficient or universally futile.
The more responsible conclusion is narrower: robustness depends on attacker capability, watermark family, deployment format, and the extent to which attacked outputs remain useful after manipulation.


\section{Generative AI Usage}
Generative AI assistance was used for manuscript revision, code inspection and changes, analysis-script development, experiment-artifact analysis, and source--output content inspection.

\section{Implementation Details}
\label{sec:appendix_impl_details}

This appendix summarizes the concrete implementation choices behind \sysname.

\subsection{Default Benchmark Configuration}
\label{ssec:appendix_default_benchmark}

\mypara{Benchmark Matrix}
Unless otherwise noted, the main removal benchmark uses three instruction-tuned victim families: Llama 3.1 8B Instruct, Qwen2.5 7B Instruct, and Mistral 7B Instruct v0.3.
We evaluate on C4, Dolly-15K, and MMW BookReport, and randomly sample 100 prompts per dataset.
For C4, we prepend an instruction-style continuation prefix so that continuation-style data better matches the chat-oriented victims used in the benchmark.

\mypara{Watermark Families}
The reported benchmark covers six watermark families: \KGW, \Unigram, \SIR, \UW, \DIP, and \SynthID.
For \KGW stealing-based analysis, we additionally expose an explicit four-setting grid through repo configs with $h \in \{1,3\}$, $\gamma \in \{0.25, 0.5\}$, and $\delta=2.0$.
Other watermark families are run through the unified framework interface and compared under the same downstream reporting protocol.

\mypara{Detection And Quality Reporting}
Detector calibration targets an FPR of $1\%$, with decisions taken at the recorded run-specific thresholds.
We then report $\mathrm{TPR}_{\text{before}}$, $\mathrm{TPR}_{\text{after}}$, and $\mathrm{ASR}$.
For utility-aware reporting, we use Prometheus as the default gating metric and report $QSR@2$, $QSR@3$, and $QSR@4$, with $QSR$ in the main text referring to $QSR@3$ unless otherwise noted.
The paper-facing quality matrices also report BERTScore and PSP as supporting semantic diagnostics.

\mypara{Sample Universe}
Each attack block uses its completed-output population and generated source texts.
TPR$_b$ uses available before-detection results; ASR uses before-positive samples with both detector states, and QSR uses all completed outputs (\autoref{ssec:metrics}).
The result data record the numerator and denominator of each rate and the count of missing predictions.

\mypara{Attack Defaults}
The attack defaults below are repo settings used to instantiate benchmark runs.
For the general-scrubbing benchmark, \DIPPER uses lexical diversity $60$, order diversity $0$, sentence interval $1$, and a bf16 paraphraser backend; \LLMP uses the rewrite prompt template with max new tokens $256$, temperature $0.7$, and sampling enabled; \SIRA uses self-information threshold $30$, max new tokens $256$, and deterministic decoding, following the official implementation's percentile-style blanking semantics; \RW uses 200 total steps, 40 target valid steps, span length $6$, backtrack patience $40$, reward-model tie threshold $0.02$, and T5 sampling with top-$p=0.8$, top-$k=40$, and temperature $1.0$; and \SA uses $\alpha=1.0$, top-$k=10$, 200 confidence-normalization samples, 100 confidence bins, and max new tokens $200$.
For protocol-separated stealing-based analysis, \MIP uses Gurobi with a 1800s time limit, \MIP gap $0.05$, AS2 attack mode, 2000 stealing samples, max edit ratio $0.3$, and the greedy removal backend by default.
\SCTS uses context size $1$, at most five candidates, query top-$k=10$, step size $2$, max substitution ratio $0.1$, and chi-square threshold $0.01$.
\DEMARK is run against a \KGW-style target with context width $1$, query batch size $32$, sampled token budget $20$, predicted/actual $\delta=2.0$, and query caching enabled.
\JSV uses a query budget of 30000, batch size $10$, previous-context width $3$, prebuilt corpus reuse, and downstream scrub/spoof deltas of $\pm 7.5$.
\Bfour runs in prebuilt-corpus mode with automatic corpus discovery, OPT-IML-1.3B as the default proxy-reference model, contrastive top-$k=10$, max new tokens $256$, and fixed $\lambda=0.5$.

\mypara{Capability And Cost Transparency}
\autoref{tab:appendix_operational_profile} reports attack execution time summed from per-sample durations.
Preparation, corpus acquisition, and offline recovery are excluded as specified for each method.

\begin{table*}[t]
\centering
\small
\caption{Attack execution time across four KGW settings per method. Ranges sum per-sample durations within each run; excluded preparation stages are listed separately.}
\label{tab:appendix_operational_profile}
\setlength{\tabcolsep}{8pt}
\renewcommand{\arraystretch}{0.85}
\begin{tabular}{@{}llrrl@{}}
\toprule
Method & Objective & Runs & Execution time (min) & Excluded stages \\
\midrule
\Bfour & Scrub & 4 & 16.6--18.9 & Preparation and corpus acquisition \\
\DEMARK & Scrub & 4 & 666.0--690.2 & Preparation and prior cache construction \\
\JSV & Scrub & 4 & 6.0--13.6 & Preparation and corpus acquisition \\
\MIP & Scrub & 4 & 0.09--0.15 & Offline recovery and corpus acquisition \\
\SCTS & Scrub & 4 & 2384.7--2731.0 & Preparation outside per-sample calls \\
\DEMARKSp & Spoof & 4 & 673.7--701.6 & Preparation and prior cache construction \\
\JSVSp & Spoof & 4 & 9.3--32.3 & Preparation and corpus acquisition \\
\bottomrule
\end{tabular}

\vspace{0.3em}
\footnotesize
Each run records 100 outputs except two JSV scrub runs with 99.
Scoring performed after attack execution is excluded.
\end{table*}

\mypara{Cost Drivers}
MIP's short online editing time does not include its offline recovery stage.
DE-MARK's spoof rows record about 20,000 token queries, while JSV's 30,000 figure describes a prebuilt corpus; these quantities are not interchangeable.

\subsection{Implementation Notes}
\label{ssec:appendix_impl_notes}

\mypara{Unified Framework Integration}
We integrate watermarking methods through the MarkLLM-based interface layer and adapt attack implementations to a shared pipeline for data loading, generation, attack execution, detection, quality analysis, and result persistence.

\mypara{Direct Removal Versus Stealing-Based Attacks}
We intentionally separate general scrubbers from stealing-based attacks in the paper's reporting.
General scrubbers take watermarked text and optional side information as input.
Stealing-based attacks first recover a reusable artifact such as a greenlist, token-color map, count table, or proxy distribution.
We therefore do not collapse them into a single headline leaderboard.

\mypara{Local Models And Fixed Checkpoints}
All experiments use local checkpoints rather than floating remote model identifiers.
This avoids silent version drift during a long-running benchmark campaign and keeps attack-side helper models reproducible under the same local environment.

\mypara{Prebuilt Stealing Corpora}
For stealing-based attacks that depend on large collections of watermarked and natural continuations, the benchmark defaults to canonical prebuilt corpora.
These corpora are generated with vLLM V0 rather than V1 because, at the time of writing, the V1 path did not support the custom logits-processor interface required by our watermark injection stack.

\mypara{Attack Adaptations}
SA uses the unified edit pipeline, RW includes prompt-aware cleanup and local quality control, and SIRA's proxy and rewriting roles share the framework interface.
These adaptations connect the attack implementations to the same generation, detection, and reporting components.

\section{Additional Results}
\label{app:additional_results}

\subsection{Family-Level Aggregation for RQ2}
\label{app:rq2_family_aggregate}

RQ2 in the main text summarizes family-conditioned QSR patterns with compact figures.
The detailed evidence is the set of appendix tables below.
\autoref{tab:spoof-feasibility-kgw} reports the limited \KGW spoofing feasibility slice used in RQ5.
\autoref{tab:final-promqsr-c4-detect} and \autoref{tab:final-promqsr-c4-quality} report the canonical C4 detection/QSR and attacked-text quality matrices.
\autoref{tab:final-promqsr-dolly-detect} and \autoref{tab:final-promqsr-dolly-quality} provide the same matrices for Dolly-15K.
\autoref{tab:final-promqsr-bookreport-detect} and \autoref{tab:final-promqsr-bookreport-quality} provide the same matrices for MMW BookReport.
These tables provide the cell-level values behind the main RQ figures.

\subsection{Full General-Scrubbing Matrices}
\label{app:general_scrubbing_matrices}

The main text reports compact C4 views for readability and protocol control.
This appendix reports the full watermark--victim-model--attack matrices for the canonical C4 slice and two external-validity slices.
Detection tables report pre-attack TPR, raw ASR, and QSR@3.
The quality tables score attacked text with BERTScore, PSP, and Prometheus.
The quality tables report the original run-level automatic measurements.
Original per-record BERTScore and PSP measurements are available for 36 cells.
For the remaining 225 cells, supplementary scoring in September 2026 evaluates the same 22,497 text pairs; all 450 cell-mean differences from the original scores are below 0.001.
The supplementary scores use full-text fields, including a shared prompt when present, and are reported separately in the result data.
\autoref{app:content_validity} gives the output-screened QSR analysis.
Thirteen Dolly--SIR cells have no observable removal opportunities; their ASR is marked ``--'' and excluded from ASR averages.
Unless otherwise noted, QSR uses Prometheus threshold 3.
``N/A'' denotes combinations excluded by the benchmark protocol.
Bold in detection tables marks the row-wise best QSR@3.
Bold in quality tables marks the row-wise best Prometheus score.

\subsection{Content Validity of Attack Outputs}
\label{app:content_validity}

We screen all 26,096 outputs across the 261 general-scrubbing cells and five attack implementations, with individual source--output inspection of 251 pairs.
Empty outputs and outputs containing only punctuation are invalid.
Repetition, unrelated responses, and loss of core content require inspection against the source and task; length alone does not establish failure.
An invalid output contributes zero to quality-gated success at every gate and remains in the original denominator.
The separate screening gate leaves the measured similarity scores and detector states unchanged.

Deterministic checks identify 2 empty and 17 punctuation-only outputs.
Source--output inspection assisted by Codex identifies 235 further invalid observations, giving 254 in total.
Of the 251 inspected pairs, 16 have no established categorical failure; 25,826 other outputs receive only the mechanical screen.
Short verse retaining its core imagery is not rejected for length alone.
Some sources are themselves degraded, so output invalidity does not by itself establish attack-caused degradation.

Let $I_i$ indicate a confirmed-invalid output, $R_i$ an observed detector removal, and $P_i$ the Prometheus score.
The additional descriptive statistic is
\[
  \mathrm{QSR}^{\mathrm{screen}}_{\tau}
  = \frac{1}{|C|}
    \sum_{i\in C}
    \indicator[R_i \land P_i\ge\tau \land I_i=0].
\]
Here $I_i=0$ means that invalidity has not been established, rather than a positive content-validity judgment.

This check removes 20 successes across 20 cells at gate 3, 32 at gate 2, and 5 at gate 4.
The table summarizes all dataset--attack groups whose gate-3 values change.
The denominator and treatment of missing detector states follow \autoref{ssec:metrics}.

\begin{center}
\small
\begin{tabular}{llrrr}
\toprule
Dataset & Attack & Removed & QSR@3 & Screened \\
\midrule
BookReport & DIPPER & 3 & 30.22 & 30.06 \\
BookReport & RW & 2 & 17.33 & 17.22 \\
C4 & DIPPER & 7 & 58.82 & 58.43 \\
C4 & RW & 2 & 36.06 & 35.94 \\
Dolly & DIPPER & 1 & 37.72 & 37.67 \\
Dolly & RW & 5 & 23.17 & 22.89 \\
\bottomrule
\end{tabular}
\end{center}
Removed counts observations; the last two columns are cell-macro QSR@3 percentages, each over 18 cells.
Unlisted groups retain their gate-3 values.
LLMP remains the leader of the dataset-level cell macro-averages under this check.

\subsection{Spoofing Feasibility}
\label{app:spoofing_feasibility}

\autoref{tab:spoof-feasibility-kgw} reports the limited \KGW spoofing feasibility slice used in RQ5.
Natural AR is the detector accept rate on ordinary unwatermarked generations.
QSSR@3 counts spoofing success only when the forged text is detected as watermarked and reaches Prometheus $\ge 3$ against the paired watermarked output.
All eight rows have 100 outputs and complete spoof-specific Prometheus scores.
The resource columns distinguish token probes for DE-MARK from prebuilt-corpus size for JSV; execution time sums per-sample attack durations.

\begin{table*}[t]
\centering
\small
\caption{Spoofing feasibility under \KGW settings on C4 with Llama-3.1-8B-Instruct.}
\label{tab:spoof-feasibility-kgw}
\setlength{\tabcolsep}{4pt}
\begin{tabular}{lcccccccc}
\toprule
\KGW & Genuine TPR$\uparrow$ & Natural AR$\downarrow$ & Method & SSR$\uparrow$ & QSSR@3$\uparrow$ & Prom$\uparrow$ & Probes / corpus & Exec. time \\
\midrule
$h{=}1,\gamma{=}0.25$ & 0.80 & 0.00 & \DEMARKSp & 1.00 & 0.47 & 2.51 & 20k token & 701.6m \\
$h{=}1,\gamma{=}0.25$ & 0.74 & 0.00 & \JSVSp & 0.00 & 0.00 & 2.02 & 30k corpus & 9.3m \\
$h{=}1,\gamma{=}0.5$ & 0.56 & 0.00 & \DEMARKSp & 1.00 & 0.64 & 2.91 & 20k token & 688.2m \\
$h{=}1,\gamma{=}0.5$ & 0.64 & 0.00 & \JSVSp & 0.00 & 0.00 & 1.95 & 30k corpus & 32.1m \\
$h{=}3,\gamma{=}0.25$ & 0.71 & 0.00 & \DEMARKSp & 1.00 & 0.55 & 2.80 & 20k token & 674.0m \\
$h{=}3,\gamma{=}0.25$ & 0.75 & 0.00 & \JSVSp & 0.01 & 0.00 & 2.05 & 30k corpus & 32.3m \\
$h{=}3,\gamma{=}0.5$ & 0.59 & 0.00 & \DEMARKSp & 0.98 & 0.56 & 2.98 & 20k token & 673.7m \\
$h{=}3,\gamma{=}0.5$ & 0.67 & 0.00 & \JSVSp & 0.01 & 0.00 & 1.85 & 30k corpus & 9.4m \\
\bottomrule
\end{tabular}
\end{table*}

\begin{table*}[tp]
\centering
\footnotesize
\caption{Full C4 general-scrubbing detection and QSR matrix.}
\label{tab:final-promqsr-c4-detect}
\setlength{\tabcolsep}{2.2pt}
\renewcommand{\arraystretch}{1.0}
\begin{tabular}{ll|ccc|ccc|ccc|ccc|ccc}
\toprule
\multirow{2}{*}{Watermark} & \multirow{2}{*}{Model} & \multicolumn{3}{c|}{\DIPPER} & \multicolumn{3}{c|}{\LLMP} & \multicolumn{3}{c|}{\SIRA} & \multicolumn{3}{c|}{\RW} & \multicolumn{3}{c}{\SA} \\
\cmidrule(lr){3-5}\cmidrule(lr){6-8}\cmidrule(lr){9-11}\cmidrule(lr){12-14}\cmidrule(lr){15-17}
& & TPR$_b\uparrow$ & ASR$\uparrow$ & QSR@3$\uparrow$ & TPR$_b\uparrow$ & ASR$\uparrow$ & QSR@3$\uparrow$ & TPR$_b\uparrow$ & ASR$\uparrow$ & QSR@3$\uparrow$ & TPR$_b\uparrow$ & ASR$\uparrow$ & QSR@3$\uparrow$ & TPR$_b\uparrow$ & ASR$\uparrow$ & QSR@3$\uparrow$ \\
\midrule
\multirow{3}{*}{\KGW} & Llama & 0.93 & 0.78 & \textbf{0.63} & 0.92 & 0.67 & 0.57 & 0.93 & 0.82 & 0.42 & 0.96 & 0.61 & 0.35 & 1.00 & 0.46 & 0.26 \\
 & Qwen & 0.96 & 0.75 & 0.61 & 1.00 & 0.88 & 0.72 & 0.96 & 0.91 & \textbf{0.74} & 0.99 & 0.62 & 0.26 & 0.95 & 0.68 & 0.33 \\
 & Mistral & 0.94 & 0.88 & 0.71 & 0.97 & 0.82 & \textbf{0.78} & 0.94 & 0.89 & 0.44 & 0.95 & 0.81 & 0.39 & N/A & N/A & N/A \\
\midrule
\multirow{3}{*}{\Unigram} & Llama & 0.72 & 0.75 & 0.46 & 0.77 & 0.75 & \textbf{0.53} & 0.72 & 0.96 & 0.43 & 0.73 & 0.88 & 0.30 & 0.73 & 0.49 & 0.19 \\
 & Qwen & 0.81 & 0.85 & 0.57 & 0.86 & 0.98 & \textbf{0.78} & 0.81 & 0.95 & 0.65 & 0.80 & 0.64 & 0.24 & 0.75 & 0.89 & 0.22 \\
 & Mistral & 0.90 & 0.60 & 0.50 & 0.81 & 0.78 & \textbf{0.58} & 0.90 & 0.88 & 0.47 & 0.91 & 0.59 & 0.31 & N/A & N/A & N/A \\
\midrule
\multirow{3}{*}{\SIR} & Llama & 0.62 & 0.62 & 0.33 & 0.55 & 0.78 & \textbf{0.40} & 0.51 & 0.84 & 0.27 & 0.49 & 0.71 & 0.23 & 0.59 & 0.76 & 0.26 \\
 & Qwen & 0.69 & 0.79 & 0.41 & 0.70 & 0.96 & \textbf{0.63} & 0.77 & 0.95 & 0.62 & 0.71 & 0.76 & 0.32 & 0.70 & 0.91 & 0.24 \\
 & Mistral & 0.91 & 0.55 & 0.43 & 0.91 & 0.69 & \textbf{0.61} & 0.91 & 0.73 & 0.38 & 0.93 & 0.80 & 0.32 & N/A & N/A & N/A \\
\midrule
\multirow{3}{*}{\UW} & Llama & 0.90 & 0.93 & 0.71 & 0.85 & 0.92 & \textbf{0.75} & 0.90 & 0.96 & 0.47 & 0.87 & 0.78 & 0.48 & 0.90 & 0.73 & 0.24 \\
 & Qwen & 0.84 & 0.99 & 0.68 & 0.83 & 1.00 & \textbf{0.76} & 0.84 & 1.00 & 0.71 & 0.75 & 0.93 & 0.37 & 0.82 & 1.00 & 0.26 \\
 & Mistral & 0.95 & 0.96 & \textbf{0.84} & 0.94 & 0.93 & 0.82 & 0.95 & 0.96 & 0.43 & 0.98 & 0.84 & 0.50 & 0.95 & 0.99 & 0.46 \\
\midrule
\multirow{3}{*}{\DIP} & Llama & 0.57 & 1.00 & 0.49 & 0.72 & 0.90 & \textbf{0.61} & 0.57 & 0.93 & 0.37 & 0.70 & 0.89 & 0.40 & 0.75 & 0.97 & 0.19 \\
 & Qwen & 0.82 & 0.94 & 0.59 & 0.81 & 0.96 & \textbf{0.72} & 0.82 & 0.96 & 0.69 & 0.78 & 0.88 & 0.34 & 0.86 & 0.99 & 0.32 \\
 & Mistral & 0.85 & 0.94 & 0.68 & 0.84 & 0.95 & \textbf{0.76} & 0.85 & 0.92 & 0.35 & 0.85 & 0.85 & 0.37 & 0.78 & 0.99 & 0.46 \\
\midrule
\multirow{3}{*}{\SynthID} & Llama & 0.95 & 0.80 & 0.67 & 0.99 & 0.89 & \textbf{0.85} & 0.95 & 0.87 & 0.43 & 0.96 & 0.86 & 0.47 & 0.96 & 0.95 & 0.22 \\
 & Qwen & 0.98 & 0.85 & 0.70 & 0.93 & 0.97 & 0.85 & 0.98 & 0.98 & \textbf{0.87} & 0.94 & 0.88 & 0.42 & 0.95 & 1.00 & 0.27 \\
 & Mistral & 0.97 & 0.69 & 0.58 & 0.99 & 0.81 & \textbf{0.76} & 0.97 & 0.74 & 0.40 & 0.96 & 0.82 & 0.42 & 0.97 & 0.91 & 0.38 \\
\bottomrule
\end{tabular}
\end{table*}

\begin{table*}[tp]
\centering
\footnotesize
\caption{Full C4 general-scrubbing attacked-text quality matrix.}
\label{tab:final-promqsr-c4-quality}
\setlength{\tabcolsep}{3.5pt}
\renewcommand{\arraystretch}{1.0}
\begin{tabular}{ll|ccc|ccc|ccc|ccc|ccc}
\toprule
\multirow{2}{*}{Watermark} & \multirow{2}{*}{Model} & \multicolumn{3}{c|}{\DIPPER} & \multicolumn{3}{c|}{\LLMP} & \multicolumn{3}{c|}{\SIRA} & \multicolumn{3}{c|}{\RW} & \multicolumn{3}{c}{\SA} \\
\cmidrule(lr){3-5}\cmidrule(lr){6-8}\cmidrule(lr){9-11}\cmidrule(lr){12-14}\cmidrule(lr){15-17}
& & BERT$\uparrow$ & PSP$\uparrow$ & Prom$\uparrow$ & BERT$\uparrow$ & PSP$\uparrow$ & Prom$\uparrow$ & BERT$\uparrow$ & PSP$\uparrow$ & Prom$\uparrow$ & BERT$\uparrow$ & PSP$\uparrow$ & Prom$\uparrow$ & BERT$\uparrow$ & PSP$\uparrow$ & Prom$\uparrow$ \\
\midrule
\multirow{3}{*}{\KGW} & Llama & 0.929 & 0.835 & 3.70 & 0.932 & 0.844 & \textbf{4.18} & 0.910 & 0.781 & 3.09 & 0.917 & 0.819 & 3.06 & 0.877 & 0.655 & 2.61 \\
 & Qwen & 0.894 & 0.807 & 3.55 & 0.918 & 0.838 & 3.67 & 0.901 & 0.819 & \textbf{3.93} & 0.904 & 0.834 & 2.78 & 0.850 & 0.614 & 2.56 \\
 & Mistral & 0.921 & 0.821 & 3.57 & 0.938 & 0.873 & \textbf{4.45} & 0.897 & 0.749 & 2.79 & 0.899 & 0.797 & 2.70 & N/A & N/A & N/A \\
\midrule
\multirow{3}{*}{\Unigram} & Llama & 0.921 & 0.814 & 3.57 & 0.941 & 0.879 & \textbf{4.26} & 0.900 & 0.764 & 2.83 & 0.907 & 0.811 & 2.88 & 0.877 & 0.643 & 2.65 \\
 & Qwen & 0.911 & 0.837 & 3.63 & 0.922 & 0.863 & \textbf{4.19} & 0.911 & 0.836 & 4.09 & 0.908 & 0.824 & 2.71 & 0.855 & 0.604 & 2.18 \\
 & Mistral & 0.932 & 0.854 & 3.89 & 0.934 & 0.863 & \textbf{4.42} & 0.906 & 0.775 & 3.02 & 0.909 & 0.811 & 2.92 & N/A & N/A & N/A \\
\midrule
\multirow{3}{*}{\SIR} & Llama & 0.929 & 0.822 & 3.58 & 0.936 & 0.846 & \textbf{4.11} & 0.910 & 0.765 & 3.16 & 0.915 & 0.799 & 3.06 & 0.880 & 0.643 & 2.65 \\
 & Qwen & 0.900 & 0.822 & 3.46 & 0.918 & 0.858 & \textbf{4.47} & 0.905 & 0.832 & 4.17 & 0.908 & 0.849 & 3.03 & 0.851 & 0.593 & 2.45 \\
 & Mistral & 0.924 & 0.833 & 3.67 & 0.935 & 0.870 & \textbf{4.35} & 0.897 & 0.768 & 2.92 & 0.902 & 0.803 & 2.68 & N/A & N/A & N/A \\
\midrule
\multirow{3}{*}{\UW} & Llama & 0.928 & 0.831 & 3.68 & 0.941 & 0.871 & \textbf{4.53} & 0.904 & 0.770 & 2.80 & 0.915 & 0.834 & 3.17 & 0.866 & 0.754 & 2.05 \\
 & Qwen & 0.909 & 0.836 & 3.61 & 0.923 & 0.872 & \textbf{4.17} & 0.909 & 0.839 & 4.01 & 0.906 & 0.836 & 2.90 & 0.863 & 0.789 & 2.09 \\
 & Mistral & 0.928 & 0.844 & 3.90 & 0.937 & 0.875 & \textbf{4.13} & 0.901 & 0.765 & 2.72 & 0.906 & 0.815 & 2.86 & 0.864 & 0.761 & 2.56 \\
\midrule
\multirow{3}{*}{\DIP} & Llama & 0.928 & 0.832 & 3.54 & 0.940 & 0.872 & \textbf{4.24} & 0.906 & 0.771 & 3.21 & 0.919 & 0.834 & 3.07 & 0.864 & 0.719 & 1.83 \\
 & Qwen & 0.905 & 0.833 & 3.45 & 0.920 & 0.868 & \textbf{4.27} & 0.910 & 0.840 & 4.18 & 0.909 & 0.849 & 2.78 & 0.865 & 0.787 & 2.22 \\
 & Mistral & 0.925 & 0.831 & 3.59 & 0.940 & 0.881 & \textbf{4.41} & 0.898 & 0.749 & 2.58 & 0.908 & 0.813 & 2.89 & 0.862 & 0.770 & 2.75 \\
\midrule
\multirow{3}{*}{\SynthID} & Llama & 0.929 & 0.838 & 3.63 & 0.940 & 0.872 & \textbf{4.38} & 0.904 & 0.771 & 2.72 & 0.914 & 0.812 & 2.94 & 0.867 & 0.735 & 1.82 \\
 & Qwen & 0.903 & 0.826 & 3.69 & 0.917 & 0.862 & \textbf{4.25} & 0.905 & 0.833 & 4.05 & 0.906 & 0.840 & 2.71 & 0.862 & 0.791 & 1.97 \\
 & Mistral & 0.929 & 0.850 & 3.85 & 0.939 & 0.872 & \textbf{4.27} & 0.896 & 0.751 & 2.80 & 0.908 & 0.819 & 2.89 & 0.862 & 0.770 & 2.29 \\
\bottomrule
\end{tabular}
\end{table*}

\clearpage

\begin{table*}[tp]
\centering
\footnotesize
\caption{Full Dolly-15K general-scrubbing detection and QSR matrix.}
\label{tab:final-promqsr-dolly-detect}
\setlength{\tabcolsep}{2.2pt}
\renewcommand{\arraystretch}{1.0}
\begin{tabular}{ll|ccc|ccc|ccc|ccc|ccc}
\toprule
\multirow{2}{*}{Watermark} & \multirow{2}{*}{Model} & \multicolumn{3}{c|}{\DIPPER} & \multicolumn{3}{c|}{\LLMP} & \multicolumn{3}{c|}{\SIRA} & \multicolumn{3}{c|}{\RW} & \multicolumn{3}{c}{\SA} \\
\cmidrule(lr){3-5}\cmidrule(lr){6-8}\cmidrule(lr){9-11}\cmidrule(lr){12-14}\cmidrule(lr){15-17}
& & TPR$_b\uparrow$ & ASR$\uparrow$ & QSR@3$\uparrow$ & TPR$_b\uparrow$ & ASR$\uparrow$ & QSR@3$\uparrow$ & TPR$_b\uparrow$ & ASR$\uparrow$ & QSR@3$\uparrow$ & TPR$_b\uparrow$ & ASR$\uparrow$ & QSR@3$\uparrow$ & TPR$_b\uparrow$ & ASR$\uparrow$ & QSR@3$\uparrow$ \\
\midrule
\multirow{3}{*}{\KGW} & Llama & 0.77 & 0.87 & 0.44 & 0.81 & 0.83 & \textbf{0.63} & 0.77 & 0.83 & 0.45 & 0.82 & 0.67 & 0.27 & 0.77 & 0.58 & 0.21 \\
 & Qwen & 0.85 & 0.88 & 0.58 & 0.90 & 0.99 & \textbf{0.83} & 0.85 & 0.98 & 0.67 & 0.85 & 0.68 & 0.30 & 0.87 & 0.86 & 0.38 \\
 & Mistral & 0.85 & 0.89 & 0.56 & 0.87 & 0.86 & \textbf{0.70} & 0.85 & 0.88 & 0.48 & 0.80 & 0.82 & 0.28 & N/A & N/A & N/A \\
\midrule
\multirow{3}{*}{\Unigram} & Llama & 0.39 & 0.82 & 0.25 & 0.42 & 0.90 & \textbf{0.35} & 0.39 & 0.95 & 0.18 & 0.42 & 0.83 & 0.19 & 0.36 & 0.56 & 0.12 \\
 & Qwen & 0.92 & 0.50 & 0.31 & 0.93 & 0.83 & \textbf{0.73} & 0.92 & 0.82 & 0.61 & 0.91 & 0.35 & 0.19 & 0.93 & 0.54 & 0.31 \\
 & Mistral & 0.56 & 0.77 & 0.31 & 0.48 & 0.88 & \textbf{0.41} & 0.56 & 0.82 & 0.30 & 0.51 & 0.61 & 0.13 & N/A & N/A & N/A \\
\midrule
\multirow{3}{*}{\SIR} & Llama & 0.00 & -- & \textbf{0.00} & 0.00 & -- & \textbf{0.00} & 0.00 & -- & \textbf{0.00} & 0.01 & 1.00 & \textbf{0.00} & 0.00 & -- & \textbf{0.00} \\
 & Qwen & 0.00 & -- & \textbf{0.00} & 0.00 & -- & \textbf{0.00} & 0.00 & -- & \textbf{0.00} & 0.00 & -- & \textbf{0.00} & 0.00 & -- & \textbf{0.00} \\
 & Mistral & 0.00 & -- & \textbf{0.00} & 0.00 & -- & \textbf{0.00} & 0.00 & -- & \textbf{0.00} & 0.00 & -- & \textbf{0.00} & N/A & N/A & N/A \\
\midrule
\multirow{3}{*}{\UW} & Llama & 0.71 & 1.00 & 0.57 & 0.70 & 1.00 & \textbf{0.67} & 0.71 & 0.97 & 0.49 & 0.67 & 0.97 & 0.40 & 0.72 & 1.00 & 0.23 \\
 & Qwen & 0.90 & 1.00 & 0.74 & 0.89 & 1.00 & \textbf{0.82} & 0.90 & 1.00 & 0.75 & 0.94 & 0.84 & 0.48 & 0.91 & 1.00 & 0.55 \\
 & Mistral & 0.59 & 1.00 & 0.47 & 0.70 & 1.00 & \textbf{0.65} & 0.59 & 1.00 & 0.41 & 0.70 & 0.91 & 0.30 & 0.64 & 1.00 & 0.38 \\
\midrule
\multirow{3}{*}{\DIP} & Llama & 0.50 & 0.98 & 0.36 & 0.48 & 1.00 & \textbf{0.47} & 0.50 & 0.80 & 0.21 & 0.51 & 0.86 & 0.29 & 0.47 & 0.91 & 0.10 \\
 & Qwen & 0.67 & 0.96 & 0.44 & 0.60 & 1.00 & \textbf{0.58} & 0.67 & 0.97 & 0.47 & 0.63 & 0.87 & 0.26 & 0.65 & 1.00 & 0.37 \\
 & Mistral & 0.85 & 0.93 & 0.59 & 0.78 & 0.94 & \textbf{0.64} & 0.72 & 0.92 & 0.37 & 0.71 & 0.90 & 0.34 & 0.78 & 0.96 & 0.48 \\
\midrule
\multirow{3}{*}{\SynthID} & Llama & 0.49 & 0.96 & 0.39 & 0.46 & 0.98 & \textbf{0.44} & 0.49 & 0.94 & 0.28 & 0.51 & 0.86 & 0.22 & 0.50 & 1.00 & 0.17 \\
 & Qwen & 0.61 & 0.93 & 0.40 & 0.67 & 1.00 & \textbf{0.60} & 0.61 & 1.00 & 0.48 & 0.63 & 0.87 & 0.37 & 0.58 & 1.00 & 0.28 \\
 & Mistral & 0.51 & 0.98 & 0.38 & 0.47 & 0.98 & \textbf{0.44} & 0.51 & 0.98 & 0.35 & 0.42 & 0.88 & 0.15 & 0.53 & 1.00 & 0.35 \\
\bottomrule
\end{tabular}
\end{table*}

\begin{table*}[tp]
\centering
\footnotesize
\caption{Full Dolly-15K general-scrubbing attacked-text quality matrix.}
\label{tab:final-promqsr-dolly-quality}
\setlength{\tabcolsep}{3.5pt}
\renewcommand{\arraystretch}{1.0}
\begin{tabular}{ll|ccc|ccc|ccc|ccc|ccc}
\toprule
\multirow{2}{*}{Watermark} & \multirow{2}{*}{Model} & \multicolumn{3}{c|}{\DIPPER} & \multicolumn{3}{c|}{\LLMP} & \multicolumn{3}{c|}{\SIRA} & \multicolumn{3}{c|}{\RW} & \multicolumn{3}{c}{\SA} \\
\cmidrule(lr){3-5}\cmidrule(lr){6-8}\cmidrule(lr){9-11}\cmidrule(lr){12-14}\cmidrule(lr){15-17}
& & BERT$\uparrow$ & PSP$\uparrow$ & Prom$\uparrow$ & BERT$\uparrow$ & PSP$\uparrow$ & Prom$\uparrow$ & BERT$\uparrow$ & PSP$\uparrow$ & Prom$\uparrow$ & BERT$\uparrow$ & PSP$\uparrow$ & Prom$\uparrow$ & BERT$\uparrow$ & PSP$\uparrow$ & Prom$\uparrow$ \\
\midrule
\multirow{3}{*}{\KGW} & Llama & 0.910 & 0.818 & 3.10 & 0.939 & 0.882 & \textbf{4.47} & 0.897 & 0.758 & 3.49 & 0.915 & 0.850 & 3.16 & 0.872 & 0.677 & 2.68 \\
 & Qwen & 0.905 & 0.801 & 3.18 & 0.928 & 0.856 & \textbf{4.32} & 0.894 & 0.778 & 3.80 & 0.910 & 0.854 & 2.99 & 0.865 & 0.651 & 2.54 \\
 & Mistral & 0.914 & 0.826 & 3.19 & 0.932 & 0.861 & \textbf{4.43} & 0.895 & 0.755 & 3.32 & 0.894 & 0.772 & 2.74 & N/A & N/A & N/A \\
\midrule
\multirow{3}{*}{\Unigram} & Llama & 0.910 & 0.810 & 3.24 & 0.938 & 0.869 & \textbf{4.41} & 0.893 & 0.756 & 3.15 & 0.921 & 0.868 & 3.30 & 0.871 & 0.683 & 2.99 \\
 & Qwen & 0.905 & 0.810 & 3.25 & 0.926 & 0.850 & \textbf{4.40} & 0.899 & 0.786 & 3.71 & 0.911 & 0.844 & 3.09 & 0.863 & 0.652 & 2.70 \\
 & Mistral & 0.919 & 0.840 & 3.39 & 0.928 & 0.856 & \textbf{4.27} & 0.897 & 0.755 & 3.12 & 0.901 & 0.792 & 2.87 & N/A & N/A & N/A \\
\midrule
\multirow{3}{*}{\SIR} & Llama & 0.910 & 0.813 & 3.10 & 0.938 & 0.875 & \textbf{4.48} & 0.898 & 0.766 & 3.34 & 0.912 & 0.856 & 2.82 & 0.870 & 0.666 & 2.99 \\
 & Qwen & 0.898 & 0.783 & 3.15 & 0.924 & 0.853 & \textbf{4.44} & 0.897 & 0.782 & 3.76 & 0.919 & 0.870 & 3.37 & 0.862 & 0.641 & 2.83 \\
 & Mistral & 0.915 & 0.832 & 3.21 & 0.929 & 0.857 & \textbf{4.41} & 0.890 & 0.734 & 3.36 & 0.902 & 0.802 & 2.81 & N/A & N/A & N/A \\
\midrule
\multirow{3}{*}{\UW} & Llama & 0.916 & 0.836 & 3.22 & 0.941 & 0.890 & \textbf{4.49} & 0.904 & 0.790 & 3.23 & 0.920 & 0.868 & 2.96 & 0.857 & 0.756 & 1.86 \\
 & Qwen & 0.909 & 0.817 & 3.35 & 0.925 & 0.851 & \textbf{4.42} & 0.895 & 0.771 & 3.73 & 0.916 & 0.867 & 3.07 & 0.889 & 0.845 & 2.86 \\
 & Mistral & 0.919 & 0.844 & 3.39 & 0.931 & 0.860 & \textbf{4.33} & 0.894 & 0.751 & 3.39 & 0.906 & 0.820 & 2.68 & 0.884 & 0.833 & 2.99 \\
\midrule
\multirow{3}{*}{\DIP} & Llama & 0.914 & 0.833 & 3.19 & 0.936 & 0.877 & \textbf{4.44} & 0.903 & 0.787 & 3.29 & 0.924 & 0.875 & 3.07 & 0.858 & 0.759 & 1.80 \\
 & Qwen & 0.903 & 0.784 & 3.19 & 0.928 & 0.856 & \textbf{4.38} & 0.897 & 0.772 & 3.80 & 0.906 & 0.831 & 2.98 & 0.889 & 0.848 & 2.63 \\
 & Mistral & 0.915 & 0.825 & 3.13 & 0.931 & 0.862 & \textbf{4.12} & 0.890 & 0.737 & 3.03 & 0.905 & 0.811 & 2.91 & 0.881 & 0.828 & 2.99 \\
\midrule
\multirow{3}{*}{\SynthID} & Llama & 0.911 & 0.808 & 3.41 & 0.938 & 0.878 & \textbf{4.36} & 0.901 & 0.781 & 3.33 & 0.917 & 0.859 & 3.00 & 0.858 & 0.765 & 2.03 \\
 & Qwen & 0.902 & 0.779 & 3.11 & 0.927 & 0.848 & \textbf{4.44} & 0.898 & 0.781 & 3.79 & 0.913 & 0.840 & 3.16 & 0.884 & 0.837 & 2.63 \\
 & Mistral & 0.920 & 0.837 & 3.33 & 0.931 & 0.859 & \textbf{4.34} & 0.896 & 0.753 & 3.25 & 0.905 & 0.808 & 2.95 & 0.882 & 0.827 & 2.90 \\
\bottomrule
\end{tabular}
\end{table*}

\clearpage

\begin{table*}[tp]
\centering
\footnotesize
\caption{Full MMW BookReport general-scrubbing detection and QSR matrix.}
\label{tab:final-promqsr-bookreport-detect}
\setlength{\tabcolsep}{2.2pt}
\renewcommand{\arraystretch}{1.0}
\begin{tabular}{ll|ccc|ccc|ccc|ccc|ccc}
\toprule
\multirow{2}{*}{Watermark} & \multirow{2}{*}{Model} & \multicolumn{3}{c|}{\DIPPER} & \multicolumn{3}{c|}{\LLMP} & \multicolumn{3}{c|}{\SIRA} & \multicolumn{3}{c|}{\RW} & \multicolumn{3}{c}{\SA} \\
\cmidrule(lr){3-5}\cmidrule(lr){6-8}\cmidrule(lr){9-11}\cmidrule(lr){12-14}\cmidrule(lr){15-17}
& & TPR$_b\uparrow$ & ASR$\uparrow$ & QSR@3$\uparrow$ & TPR$_b\uparrow$ & ASR$\uparrow$ & QSR@3$\uparrow$ & TPR$_b\uparrow$ & ASR$\uparrow$ & QSR@3$\uparrow$ & TPR$_b\uparrow$ & ASR$\uparrow$ & QSR@3$\uparrow$ & TPR$_b\uparrow$ & ASR$\uparrow$ & QSR@3$\uparrow$ \\
\midrule
\multirow{3}{*}{\KGW} & Llama & 0.90 & 0.97 & 0.40 & 0.93 & 0.69 & \textbf{0.55} & 0.90 & 0.90 & 0.38 & 0.93 & 0.76 & 0.30 & 0.94 & 0.59 & 0.24 \\
 & Qwen & 0.99 & 0.89 & 0.24 & 0.98 & 0.97 & 0.56 & 0.99 & 0.94 & \textbf{0.68} & 0.97 & 0.74 & 0.13 & 1.00 & 0.60 & 0.18 \\
 & Mistral & 0.90 & 0.99 & 0.36 & 0.89 & 0.83 & \textbf{0.66} & 0.90 & 0.98 & 0.34 & 0.95 & 0.89 & 0.10 & N/A & N/A & N/A \\
\midrule
\multirow{3}{*}{\Unigram} & Llama & 0.74 & 0.69 & 0.27 & 0.71 & 0.65 & \textbf{0.44} & 0.74 & 0.96 & 0.33 & 0.67 & 0.85 & 0.19 & 0.76 & 0.37 & 0.13 \\
 & Qwen & 0.98 & 0.69 & 0.28 & 0.98 & 0.81 & \textbf{0.69} & 0.98 & 0.72 & 0.53 & 0.93 & 0.59 & 0.16 & 0.96 & 0.48 & 0.22 \\
 & Mistral & 0.76 & 0.88 & 0.30 & 0.57 & 0.89 & \textbf{0.45} & 0.76 & 0.92 & 0.23 & 0.71 & 0.55 & 0.04 & N/A & N/A & N/A \\
\midrule
\multirow{3}{*}{\SIR} & Llama & 0.68 & 0.71 & 0.25 & 0.65 & 0.77 & \textbf{0.45} & 0.68 & 0.87 & 0.38 & 0.54 & 0.74 & 0.12 & 0.59 & 0.80 & 0.16 \\
 & Qwen & 0.90 & 0.72 & 0.26 & 0.88 & 0.98 & 0.52 & 0.90 & 0.87 & \textbf{0.53} & 0.67 & 0.76 & 0.17 & 0.59 & 0.90 & 0.22 \\
 & Mistral & 0.77 & 0.66 & 0.17 & 0.82 & 0.68 & \textbf{0.51} & 0.77 & 0.94 & 0.30 & 0.84 & 0.70 & 0.08 & N/A & N/A & N/A \\
\midrule
\multirow{3}{*}{\UW} & Llama & 0.99 & 0.98 & 0.40 & 1.00 & 0.91 & \textbf{0.83} & 0.99 & 0.96 & 0.58 & 0.99 & 0.72 & 0.28 & 0.98 & 0.32 & 0.07 \\
 & Qwen & 1.00 & 1.00 & 0.34 & 1.00 & 1.00 & \textbf{0.86} & 1.00 & 1.00 & 0.81 & 1.00 & 0.83 & 0.39 & 1.00 & 0.90 & 0.35 \\
 & Mistral & 1.00 & 0.95 & 0.36 & 1.00 & 0.97 & \textbf{0.85} & 1.00 & 0.94 & 0.34 & 1.00 & 0.88 & 0.09 & 1.00 & 0.80 & 0.45 \\
\midrule
\multirow{3}{*}{\DIP} & Llama & 0.50 & 0.98 & 0.27 & 0.44 & 1.00 & \textbf{0.39} & 0.50 & 0.96 & 0.27 & 0.34 & 1.00 & 0.08 & 0.37 & 1.00 & 0.03 \\
 & Qwen & 0.82 & 0.99 & 0.45 & 0.89 & 0.93 & 0.66 & 0.82 & 0.99 & \textbf{0.75} & 0.87 & 0.87 & 0.32 & 0.89 & 0.99 & 0.46 \\
 & Mistral & 0.68 & 1.00 & 0.26 & 0.83 & 0.95 & \textbf{0.71} & 0.69 & 1.00 & 0.24 & 0.70 & 0.96 & 0.08 & 0.80 & 1.00 & 0.27 \\
\midrule
\multirow{3}{*}{\SynthID} & Llama & 0.54 & 0.93 & 0.20 & 0.82 & 0.94 & \textbf{0.71} & 0.54 & 0.93 & 0.26 & 0.75 & 0.93 & 0.25 & 0.77 & 0.97 & 0.15 \\
 & Qwen & 0.98 & 0.83 & 0.31 & 0.99 & 0.94 & \textbf{0.74} & 0.98 & 0.92 & 0.71 & 0.98 & 0.78 & 0.26 & 1.00 & 0.87 & 0.40 \\
 & Mistral & 0.89 & 0.84 & 0.32 & 0.85 & 0.92 & \textbf{0.70} & 0.89 & 0.91 & 0.36 & 0.87 & 0.90 & 0.08 & 0.91 & 0.90 & 0.35 \\
\bottomrule
\end{tabular}
\end{table*}

\begin{table*}[tp]
\centering
\footnotesize
\caption{Full MMW BookReport general-scrubbing attacked-text quality matrix.}
\label{tab:final-promqsr-bookreport-quality}
\setlength{\tabcolsep}{3.5pt}
\renewcommand{\arraystretch}{1.0}
\begin{tabular}{ll|ccc|ccc|ccc|ccc|ccc}
\toprule
\multirow{2}{*}{Watermark} & \multirow{2}{*}{Model} & \multicolumn{3}{c|}{\DIPPER} & \multicolumn{3}{c|}{\LLMP} & \multicolumn{3}{c|}{\SIRA} & \multicolumn{3}{c|}{\RW} & \multicolumn{3}{c}{\SA} \\
\cmidrule(lr){3-5}\cmidrule(lr){6-8}\cmidrule(lr){9-11}\cmidrule(lr){12-14}\cmidrule(lr){15-17}
& & BERT$\uparrow$ & PSP$\uparrow$ & Prom$\uparrow$ & BERT$\uparrow$ & PSP$\uparrow$ & Prom$\uparrow$ & BERT$\uparrow$ & PSP$\uparrow$ & Prom$\uparrow$ & BERT$\uparrow$ & PSP$\uparrow$ & Prom$\uparrow$ & BERT$\uparrow$ & PSP$\uparrow$ & Prom$\uparrow$ \\
\midrule
\multirow{3}{*}{\KGW} & Llama & 0.900 & 0.814 & 2.26 & 0.945 & 0.913 & \textbf{4.05} & 0.874 & 0.681 & 2.56 & 0.899 & 0.796 & 2.26 & 0.881 & 0.741 & 2.43 \\
 & Qwen & 0.888 & 0.763 & 1.79 & 0.928 & 0.878 & 3.19 & 0.885 & 0.754 & \textbf{3.33} & 0.893 & 0.784 & 1.71 & 0.872 & 0.724 & 1.94 \\
 & Mistral & 0.898 & 0.786 & 2.27 & 0.932 & 0.885 & \textbf{4.15} & 0.866 & 0.650 & 2.28 & 0.856 & 0.636 & 1.35 & N/A & N/A & N/A \\
\midrule
\multirow{3}{*}{\Unigram} & Llama & 0.892 & 0.787 & 2.39 & 0.944 & 0.903 & \textbf{4.22} & 0.869 & 0.660 & 2.27 & 0.894 & 0.777 & 2.14 & 0.880 & 0.753 & 2.37 \\
 & Qwen & 0.886 & 0.750 & 2.16 & 0.929 & 0.876 & \textbf{3.94} & 0.882 & 0.743 & 3.51 & 0.903 & 0.822 & 2.10 & 0.871 & 0.723 & 2.26 \\
 & Mistral & 0.898 & 0.795 & 2.21 & 0.936 & 0.892 & \textbf{4.17} & 0.862 & 0.633 & 2.07 & 0.873 & 0.702 & 1.59 & N/A & N/A & N/A \\
\midrule
\multirow{3}{*}{\SIR} & Llama & 0.900 & 0.814 & 2.40 & 0.947 & 0.909 & \textbf{4.12} & 0.878 & 0.694 & 2.96 & 0.900 & 0.818 & 2.04 & 0.881 & 0.743 & 2.47 \\
 & Qwen & 0.892 & 0.792 & 2.09 & 0.924 & 0.863 & 3.10 & 0.884 & 0.751 & \textbf{3.20} & 0.895 & 0.789 & 2.01 & 0.875 & 0.735 & 2.15 \\
 & Mistral & 0.897 & 0.785 & 2.20 & 0.932 & 0.891 & \textbf{4.15} & 0.872 & 0.676 & 2.18 & 0.866 & 0.673 & 1.53 & N/A & N/A & N/A \\
\midrule
\multirow{3}{*}{\UW} & Llama & 0.902 & 0.830 & 2.28 & 0.947 & 0.910 & \textbf{4.10} & 0.876 & 0.702 & 2.87 & 0.899 & 0.797 & 2.26 & 0.859 & 0.802 & 1.56 \\
 & Qwen & 0.888 & 0.767 & 2.00 & 0.926 & 0.862 & \textbf{3.89} & 0.881 & 0.746 & 3.73 & 0.905 & 0.829 & 2.43 & 0.900 & 0.882 & 2.18 \\
 & Mistral & 0.900 & 0.791 & 2.16 & 0.935 & 0.889 & \textbf{4.11} & 0.867 & 0.648 & 2.11 & 0.861 & 0.662 & 1.36 & 0.890 & 0.864 & 2.62 \\
\midrule
\multirow{3}{*}{\DIP} & Llama & 0.900 & 0.812 & 2.44 & 0.944 & 0.905 & \textbf{4.12} & 0.874 & 0.688 & 2.75 & 0.898 & 0.799 & 2.10 & 0.860 & 0.782 & 1.43 \\
 & Qwen & 0.893 & 0.783 & 2.43 & 0.928 & 0.868 & 3.78 & 0.885 & 0.754 & \textbf{4.00} & 0.901 & 0.821 & 2.33 & 0.897 & 0.877 & 2.53 \\
 & Mistral & 0.900 & 0.789 & 2.19 & 0.934 & 0.895 & \textbf{4.01} & 0.869 & 0.663 & 2.08 & 0.864 & 0.656 & 1.46 & 0.882 & 0.856 & 2.18 \\
\midrule
\multirow{3}{*}{\SynthID} & Llama & 0.899 & 0.819 & 2.26 & 0.942 & 0.903 & \textbf{4.30} & 0.873 & 0.677 & 2.60 & 0.901 & 0.818 & 2.29 & 0.860 & 0.793 & 1.58 \\
 & Qwen & 0.887 & 0.771 & 2.09 & 0.925 & 0.868 & \textbf{3.75} & 0.882 & 0.748 & 3.53 & 0.898 & 0.814 & 2.23 & 0.899 & 0.881 & 2.54 \\
 & Mistral & 0.903 & 0.805 & 2.32 & 0.939 & 0.898 & \textbf{3.99} & 0.877 & 0.686 & 2.51 & 0.865 & 0.686 & 1.51 & 0.891 & 0.870 & 2.45 \\
\bottomrule
\end{tabular}
\end{table*}

\clearpage
\onecolumn
\section{Cross-Watermark Quality-Gate Sensitivity}
\label{app:synthid-sensitivity}
\mypara{SynthID configuration}
The evaluated MarkLLM SynthID-Text implementation uses iterative probability reweighting with 30 keyed binary $g$-value channels over five-token n-grams, including four preceding context tokens.
Its configuration uses a 65,536-entry sampling table, table seed 0, and a 1,024-context repetition history.
The mean detector uses the victim tokenizer without added special tokens and excludes repeated contexts and positions from EOS onward.
The stored \texttt{z\_score} field contains the mean $g$-value.

All 45 SynthID runs use dynamic detection with increasing-score decisions and target calibration FPR 0.01.
The run-specific threshold is selected from natural and watermarked score populations and applied as score $\ge\tau$; thresholds range from 0.5099 to 0.5228.
Configuration and implementation identifiers accompany the result data.

\mypara{Analysis population}
The analysis uses 180 cells: three datasets, six watermarks, Llama and Qwen, and five attacks.
These two victim models provide all five attacks for every watermark; Mistral has structural attack exclusions and remains in the full benchmark tables.
Each entry averages the Llama and Qwen cell rates with equal weight at Prometheus gates 2, 3, and 4, using the QSR denominator in \autoref{ssec:metrics}.
The 17,996 outputs include 33 with at least one missing detector prediction; these outputs remain in the QSR denominator and receive no success credit.

\begin{table}[!htbp]
\centering
\caption{Quality-gate sensitivity on common Llama/Qwen support.}
\label{tab:synthid-gate-sensitivity}
\small
\setlength{\tabcolsep}{7pt}
\begin{tabular}{llrrrrrr}
\toprule
& & \multicolumn{3}{c}{LLMP $\QSRobs@u$ (\%)} & \multicolumn{3}{c}{LLMP minus best other (pp)} \\
\cmidrule(lr){3-5}\cmidrule(lr){6-8}
Dataset & Watermark & $u=2$ & $u=3$ & $u=4$ & $u=2$ & $u=3$ & $u=4$ \\
\midrule
C4 & KGW & 68.00 & 64.50 & 56.00 & +2.00 (S) & +2.50 (D) & +6.00 (S) \\
 & Unigram & 68.00 & 65.50 & 59.50 & +8.00 (S) & +11.50 (S) & +12.50 (S) \\
 & SIR & 53.00 & 51.50 & 48.50 & +6.50 (S) & +7.00 (S) & +12.00 (S) \\
 & UW & 77.00 & 75.50 & 70.00 & +0.50 (D) & +6.00 (D) & +16.50 (D) \\
 & DIP & 70.00 & 66.50 & 60.50 & +9.00 (D) & +12.50 (D) & +15.00 (S) \\
 & \textbf{SynthID} & 87.50 & 85.00 & 77.00 & +12.11 (D) & +16.65 (D) & +24.50 (S) \\
\midrule
Dolly-15K & KGW & 75.50 & 73.00 & 69.50 & +14.50 (S) & +17.00 (S) & +21.00 (S) \\
 & Unigram & 55.50 & 54.00 & 49.50 & +12.00 (S) & +14.50 (S) & +16.00 (S) \\
 & SIR & 0.00 & 0.00 & 0.00 & +0.00 (T) & +0.00 (T) & +0.00 (T) \\
 & UW & 75.77 & 74.74 & 71.66 & +2.77 (D) & +9.24 (D) & +23.16 (S) \\
 & DIP & 53.00 & 52.50 & 48.00 & +7.00 (D) & +12.50 (D) & +18.50 (S) \\
 & \textbf{SynthID} & 52.50 & 52.00 & 50.00 & +9.50 (D) & +12.50 (D) & +18.50 (S) \\
\midrule
BookReport & KGW & 62.50 & 55.50 & 49.00 & +5.00 (S) & +2.50 (S) & +9.50 (S) \\
 & Unigram & 57.00 & 56.50 & 50.00 & +10.00 (S) & +13.50 (S) & +18.00 (S) \\
 & SIR & 52.50 & 48.50 & 42.50 & +3.50 (S) & +3.00 (S) & +7.50 (S) \\
 & UW & 87.50 & 84.50 & 74.50 & +11.00 (S) & +15.00 (S) & +22.00 (S) \\
 & DIP & 56.50 & 52.50 & 44.50 & +3.50 (S) & +1.50 (S) & +3.00 (S) \\
 & \textbf{SynthID} & 76.50 & 72.50 & 62.50 & +24.50 (S) & +24.00 (S) & +27.00 (S) \\
\bottomrule
\end{tabular}
\end{table}

\mypara{Reading the table}
Rates are percentages; differences are percentage points relative to the highest-rate non-\LLMP attack at that gate.
D denotes \DIPPER and S denotes \SIRA; T denotes a tie among all four non-\LLMP attacks.
The comparator is selected separately at each gate.
The result data provide per-cell denominators, model-specific rates, and all five attack rankings.

\end{document}